\documentclass[sigplan,screen,nonacm]{acmart}

\usepackage{amsmath}
\usepackage{wrapfig}
\usepackage{float}
\usepackage{array}
\usepackage{algorithm}
\usepackage{algpseudocode}
\algrenewcommand\algorithmicindent{1em} % Set indentation to 1em (default is 1.5em)
\usepackage{listings}
\usepackage{enumitem}
\usepackage{color}
\usepackage{xcolor}
\usepackage{soul}
\usepackage{booktabs}
\usepackage{multirow}
\usepackage{wasysym} % for circles
\usepackage{xurl}

\newcommand{\saman}[1]{}

\lstdefinestyle{mystyle}{
    backgroundcolor=\color{white},   
    commentstyle=\color{Red3},
    numberstyle=\tiny\color{gray},
    stringstyle=\color{Blue3},
    basicstyle=\footnotesize\ttfamily,
    breakatwhitespace=false,         
    breaklines=true,                 
    numbers=left,                    
    numbersep=5pt,                  
    xleftmargin=2em,
    showspaces=false,                
    showstringspaces=false,
    showtabs=false, 
    showlines=true,
    tabsize=2,
    keywordstyle=\color{purple}\bfseries,
    columns=fullflexible,
    escapeinside={(*@}{@*)},
    captionpos=b
}

\setcopyright{cc}
\setcctype{by}
\acmDOI{10.1145/3696443.3708919}
\acmYear{2025}
\copyrightyear{2025}
\acmISBN{979-8-4007-1275-3/25/03}
\acmConference[CGO '25]{Proceedings of the 23rd ACM/IEEE International Symposium on Code Generation and Optimization}{March 01--05, 2025}{Las Vegas, NV, USA}
\acmBooktitle{Proceedings of the 23rd ACM/IEEE International Symposium on Code Generation and Optimization (CGO '25), March 01--05, 2025, Las Vegas, NV, USA}
\received{2024-05-30}
\received[accepted]{2024-07-22}

\begin{document}

\title{SySTeC: A Symmetric Sparse Tensor Compiler}

\author{Radha Patel}
\affiliation{%
  \institution{MIT CSAIL}
  \city{Cambridge}
  \state{Massachusetts}
  \country{USA}}
\email{rrpatel@alum.mit.edu}

\author{Willow Ahrens}
\affiliation{%
  \institution{MIT CSAIL}
  \city{Cambridge}
  \state{Massachusetts}
  \country{USA}}
\email{willow@csail.mit.edu}

\author{Saman Amarasinghe}
\affiliation{%
  \institution{MIT CSAIL}
  \city{Cambridge}
  \state{Massachusetts}
  \country{USA}}
\email{saman@csail.mit.edu}

\renewcommand{\shortauthors}{Patel et al.}

\begin{abstract}
  Symmetric and sparse tensors arise naturally in many domains including linear algebra, statistics, physics, chemistry, and graph theory.
  Symmetric tensors are equal to their transposes, so in the $n$-dimensional case we can save up to a factor of $n!$ by avoiding redundant operations.
  Sparse tensors, on the other hand, are mostly zero, and we can save asymptotically by processing only nonzeros.
  Unfortunately, specializing for both symmetry and sparsity at the same time is uniquely challenging.
  Optimizing for symmetry requires consideration of $n!$ transpositions of a triangular kernel, which can be complex and error prone.
  Considering multiple transposed iteration orders and triangular loop bounds also complicates iteration through intricate sparse tensor formats.
  Additionally, since each combination of symmetry and sparse tensor formats requires a specialized implementation, this leads to a combinatorial number of cases.
  A compiler is needed, but existing compilers cannot take advantage of both symmetry and sparsity within the same kernel.
  In this paper, we describe the first compiler which can automatically generate symmetry-aware code for sparse or structured tensor kernels.
  We introduce a taxonomy for symmetry in tensor kernels, and show how to target each kind of symmetry.
  Our implementation demonstrates significant speedups ranging from 1.36x for SSYMV to 30.4x for a 5-dimensional MTTKRP over the non-symmetric state of the art.
\end{abstract}





\maketitle
\section{Introduction}
A symmetric tensor is a tensor that is invariant under a permutation of its indices. Tensors are often naturally symmetric because of the physical and chemical properties of substances and matter which produce symmetric interactions, structures, or reactions. Additionally, symmetry can also be induced as a mathematical consequence of how we use tensor operations (e.g. $A^TA$).

Real world tensors can also be sparse, meaning they are mostly zero or some other fill value. Special formats have been proposed to only store nonzeros and several systems, such as GraphBLAS \cite{kepner_mathematical_2016}, TACO \cite{kjolstad_tensor_2017}, or Finch \cite{ahrens_finch_2024} have been developed to efficiently perform operations on sparse tensors, but none of them can handle symmetry automatically.

There are wide-ranging applications of symmetric sparse tensors, from mathematical optimization to scientific computing. In linear algebra, the hat matrix in linear regression and the Q matrix that is a result of QR factorization are both symmetric \cite{goodall_computation_1993}. In statistics, matrices expressing covariance and other similarly commutative calculations are naturally symmetric \cite{schott_matrix_2016}. In physics and chemistry computations, the properties of quantum tensor networks and computational fluid dynamics give way to multi-dimensional symmetry \cite{orus_tensor_2019,hu_fluid_2012}. In graph theory, adjacency matrices of undirected graphs, used in algorithms like single-source shortest path and to find connected components, are also symmetric \cite{singh_role_2012}.

\begin{table}[h!]
    \centering
    \small % Reduce font size
    \setlength{\tabcolsep}{3pt} % Adjust column separation
    \renewcommand{\arraystretch}{0.8} % Adjust row separation
    \begin{tabular}{lllllll}
    \toprule
     & \rotatebox{90}{MKL} & \rotatebox{90}{TCE} & \rotatebox{90}{Cyclops} & \rotatebox{90}{sBLACs} & \rotatebox{90}{STUR} & \rotatebox{90}{SySTeC} \\
    \midrule
    Supports Dense Tensors & \CIRCLE & \CIRCLE & \CIRCLE & \LEFTcircle$^{1}$ & \CIRCLE & \CIRCLE \\
    Supports Sparse Tensors & \CIRCLE &  & \LEFTcircle$^{2}$ & \LEFTcircle$^{1,3}$ & \LEFTcircle$^{3}$ & \CIRCLE \\
    Supports Structured Tensors &  &  &  & \LEFTcircle$^{1}$ & \CIRCLE & \CIRCLE \\
    \midrule
    Supports General Einsums &  & \LEFTcircle$^{4}$ & \LEFTcircle$^{4}$ & \CIRCLE & \CIRCLE & \CIRCLE \\
    \midrule
    Optimizes Redundant Reads & \CIRCLE &  &  &  &  & \CIRCLE \\
    Optimizes Redundant Operations & \CIRCLE & \CIRCLE & \CIRCLE & \CIRCLE & \CIRCLE & \CIRCLE \\
    Optimizes Redundant Storage & \CIRCLE & \CIRCLE & \CIRCLE & \CIRCLE & \CIRCLE & \CIRCLE \\
    \bottomrule
    \end{tabular}
    \caption{Supported features: \CIRCLE = Yes, \LEFTcircle = Partially. $^{1}$ = Only static sizes, $^{2}$ = Only one sparse tensor at a time, $^{3}$ = Only symbolic patterns, $^{4}$ = Only contractions.}
    \label{tab:features}
\end{table}

Optimizing for symmetry and sparsity at once is uniquely challenging.
Symmetric optimizations require that we consider all combinatorial loop reorderings of the kernel and restrict iteration to a triangle, which can be complex and error prone.
Sparse optimizations require reformatting the data to store and process only nonzeros.
Symmetry is a property defined on the coordinates of the tensor, and sparse formats obfuscate the relationship between tensor coordinates and where elements are stored in memory.
Iterators over sparse tensor formats are often a performance bottleneck, and are especially sensitive to changes in loop ordering and loop bounds \cite{ahrens_autoscheduling_2022}.
Additionally, since each combination of symmetry and sparse tensor formats requires a specialized implementation, this leads to a combinatorial number of cases, hand-writing solutions is not feasible in the general case. Libraries such as MKL or CuBLAS only support a small subsets of symmetric sparse matrix kernels \cite{noauthor_developer_2024, noauthor_cusparse_2024}.

A compiler approach is necessary.  Though several compilers have been developed to handle symmetric tensors, none of them apply to sparse tensors, and vice versa.
Compilers like STUR \cite{ghorbani_compiling_2023} and Cyclops \cite{solomonik_sparse_2015} and sBLACs \cite{spampinato_basic_2016} all optimize for symmetric tensors, but STUR and sBLACs cannot handle unstructured sparse tensors and Cyclops cannot
handle more than one sparse tensor at a time.
These compilers accelerate symmetric kernels by avoiding redundant computation and storage, but cannot avoid redundant memory operations. In some kernels, like symmetric sparse matrix-vector multiply, we can optimize memory bandwidth by restricting iteration to the upper triangle and performing all necessary updates to the output tensor in one pass.
%
%The Cyclops Tensor Framework (CTF) reduces $N$-dimensional dense symmetric contractions to $N!$ separate triangular contractions, which are handled by looping over the triangular indices and repeatedly calling matrix multiply. \cite{solomonik2013cyclops, solomonik_sparse_2015}
%
%Though this saves compute and storage, it cannot reuse reads and writes between the separate kernels, and it does not apply to tensor kernels which are not contractions, such as MTTKRP. It only supports one sparse argument at a time.
%
%STUR uses symbolic techniques to optimize kernels based on the structure of the arguments, but only applies to static sparsity patterns. \cite{ghorbani_compiling_2023}

We aim to fill the gap by presenting a granular approach to identify and exploit symmetry in sparse tensor kernels. Our specific contributions include:

\begin{enumerate}[topsep=8pt, parsep=2pt, partopsep=0pt]
    \item To the best of our knowledge, SySTeC is the first system to automatically generate code for symmetric and sparse or otherwise structured (Triangular, Banded, Run-Length-Encoded) tensor operations.
    \item A taxonomy of symmetry in tensor kernels, and strategies to utilize each kind of symmetry. We introduce the concepts of \textit{visible} and \textit{invisible} \textit{input} and \textit{output} symmetries. Capitalizing on these saves memory bandwidth, storage, and compute by reusing reads and writes to symmetric input and filtering redundant storage and computations.
    \item We show how to extend traditional compiler optimizations to take advantage of symmetry, as well as introduce new compiler optimizations, such as \textit{simplicial lookup tables} and \textit{diagonal splitting}. Our compiler uses term rewriting to optimize redundancies, and is easily extensible to general operators beyond $+$ and $*$.
    \item We implement our compiler and evaluate it on several common tensor kernels, demonstrating speedups from 1.36x for SSYMV to 30.4x for a 5-dimensional MTTKRP with the symmetric code generated by the compiler over the naive implementation of these kernels.
\end{enumerate}

\section{Background}
In this section, we introduce the terminology and syntax that will be used throughout the rest of the paper. 

\subsection{Symmetric Tensors}
A matrix $M$ is symmetric if $M[i_1, i_2] = M[i_2, i_1]$—i.e. the entries at permutations of the indices are equivalent. We can generalize this definition for tensors \cite{comon_symmetric_2008}.

\begin{definition}[Symmetry]
\normalfont
Let T be an $n$-dimensional tensor. T is \textit{symmetric} if for all permutations $\sigma$ of $\{1, ..., n\}$, 
\[
    T[i_1, ..., i_n] = T[i_{\sigma(1)}, ..., i_{\sigma(n)}].
\]
\end{definition}

In the case of matrices, symmetry is binary: a matrix is either symmetric or it is not. However, when dealing with higher-order tensors, this definition can be expanded with the notion of \textit{partial symmetry}. A \textit{partition} $\Pi$ of a set \textit{A} is a collection of non-empty, pairwise disjoint subsets, which we will refer to as \textit{parts}, of \textit{A}, such that each element of \textit{A} belongs to exactly one subset within the collection \cite{schatz_exploiting_2013}. We denote $\pi_i$ to be the $i^{th}$ part of $\Pi$. We define partial symmetry relative to a chosen partition \cite{shi_attempt_2021}.

\begin{definition}[Partial Symmetry]
\normalfont
Let $T$ be an \linebreak[4] $n$-dimensional tensor, and let $\Pi$ be a partition of $\{1, ..., n\}$. Then $T$ is \textit{partially symmetric} if 
\[
 T[i_1, ..., i_n] = T[i_{\sigma(1)}, ..., i_{\sigma(n)}]
\]
for all permutations $\sigma$ of $\{1, ..., n\}$ which only permute elements within their parts in $\Pi$.

Then, we can denote that $T$ has $\Pi$ symmetry.
\end{definition}

Since the upper and lower triangles of a symmetric matrix are equal, our framework restricts our computations to one of the triangles to avoid redundant operations. We refer to the triangle that we choose to compute as the \textit{canonical triangle} of the tensor. We choose the upper triangle in this work.

\begin{definition}[Canonical]
\normalfont
Let tensor $T[i_1, ..., i_n]$ have symmetry $\Pi_T$. Coordinates $[i_1, ..., i_n]$ are \textit{canonical} if $i_p \leq i_q$ for any $p < q$ with $i_p$ and $i_q$ in the same part of $\Pi_T$. Otherwise, the coordinates are \textit{non-canonical}. The \textit{canonical triangle} of a tensor consists of all the canonical coordinates in the tensor.
\end{definition}

Although computations in the triangles of a tensor are repeated, computations on diagonals are not, and so diagonals must often be handled separately.

\begin{definition}[Diagonal]
\normalfont
A \textit{diagonal} of a tensor $T$ consists of all coordinates $[i_1, ..., i_n]$ where the indices in a subset $D$ of $\{i_1, ..., i_n\}$ where $|D| > 1$ are equal. 
\end{definition}

\subsection{Sparse and Structured Tensor Programming}
\begin{figure}[h]
    \noindent
\centering{\includegraphics[alt={Finch Syntax},width=\linewidth]{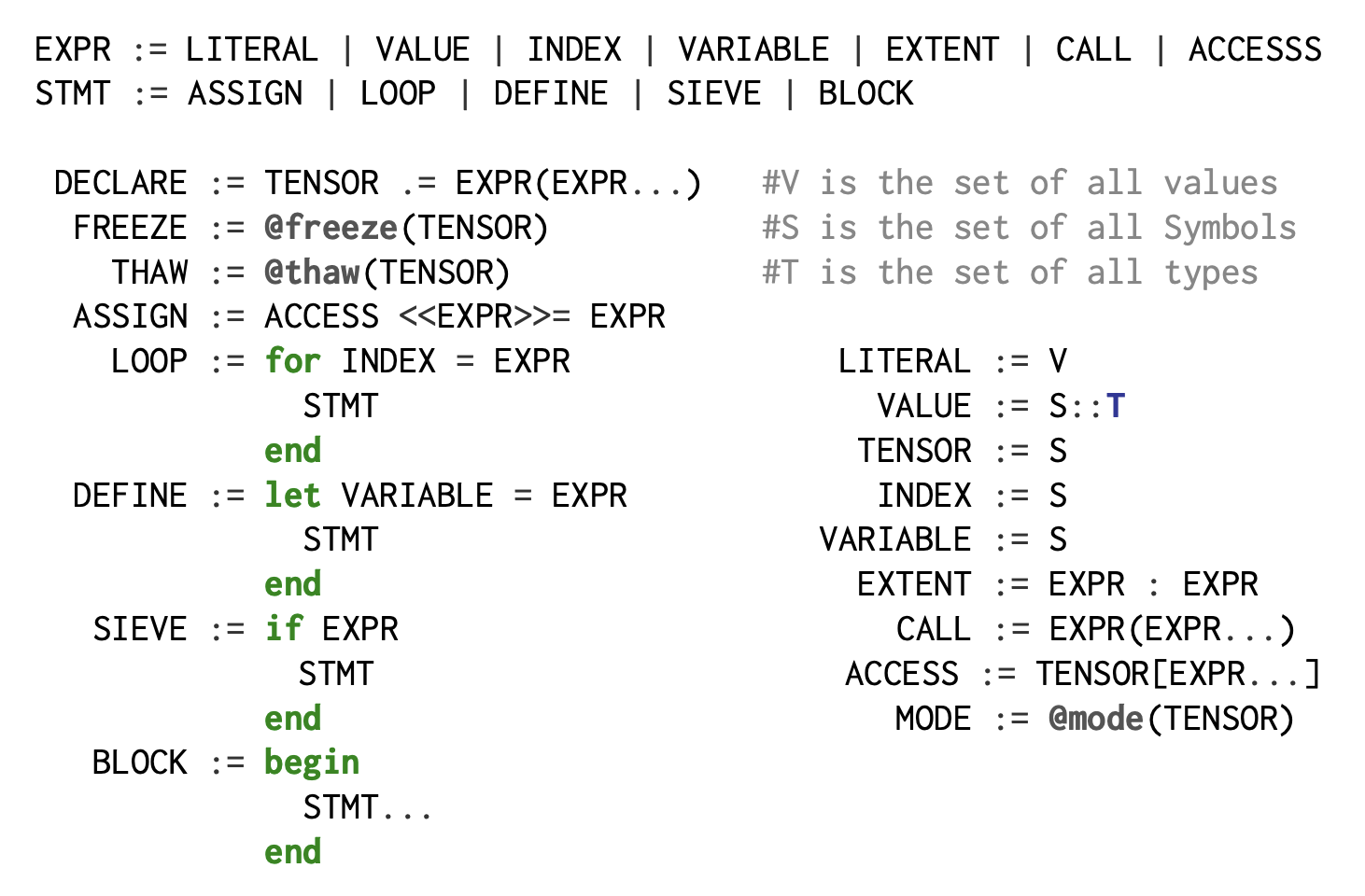}}%
\caption{Finch Syntax \cite[Figure 7]{ahrens_finch_2024}}
\label{fig:finch_syntax}
\end{figure}

We will be using the program syntax and formats from Finch, a Julia-to-Julia compiler designed for optimizing loop nests over sparse or structured (Triangular, Banded, Run-Length-Encoded) multidimensional arrays \cite{ahrens_finch_2024}. Finch supports a wide range of sparse and structured storage formats, as well as the control flow necessary to implement and execute symmetric kernels, such as conditionals and multiple outputs. Finch also simplifies the complexities of sparse data manipulation, enabling us to write \textbf{loop structures that appear dense but are compiled to be sparse} which makes it easier to focus the optimizations we apply to take advantage of symmetry. Figure \ref{fig:finch_syntax} shows the syntax of Finch.

Finch uses a hierarchical mode-by-mode fibertree description of tensor formats, where tensors are conceptualized as a vector of vectors of vectors, etc. \cite{sze_efficient_2017,chou_format_2018} This allows us to characterize each level of the tree as a separate vector type, expressing several common sparse and structured tensor formats as combinations of simple level formats. For example, CSR format is \verb|Dense(Sparse(Element(0.0)))|, or a dense vector of sparse vectors \cite{saad_iterative_2003}. The 3-dimensional CSF format is \verb|Dense(Sparse(Sparse(Element(0.0))))| \cite{smith_splatt_2015}. We refer the reader to literature on Finch for more information and examples of tensor formats \cite[Figure 6 and Table 3]{ahrens_finch_2024}.

Critically, accesses to sparse tensors in Finch syntax (e.g. \verb|x[i]|) act as iterators over 
sparse tensors, and comparisons between indices (e.g. \verb|i < j|) are lifted into loop bounds. Thus, the Finch code on left
compiles to the code on the right.

\noindent
\begin{minipage}{0.4\linewidth}
\begin{lstlisting}[linewidth=\linewidth, backgroundcolor=\color{white}, numbers=none, xleftmargin=0pt, basicstyle=\small]
x = Sparse(Element(0.0))
@finch
for i=_
    if i < 7
        s[] += x[i]       
\end{lstlisting}
\end{minipage}\hfill$\xrightarrow{Finch}$\hfill%
\begin{minipage}{0.4\linewidth}
\begin{lstlisting}[linewidth=\linewidth, backgroundcolor=\color{white}, numbers=none, basicstyle=\small, xleftmargin=0pt]
q = 1
stop = min(nnz(x), 7 - 1)
while i < stop
    i = x.idx[q]
    if i <= stop
        s += x.val[q]
        q += 1
\end{lstlisting}
\end{minipage}

\section{Techniques to Exploit Symmetry}
We categorize the symmetry that presents itself in assignments in two groups: \textbf{input symmetry}, which involves one or more input tensors being symmetric and \textbf{output symmetry}, which consists of a symmetric output tensor. Assignments can have either input or output symmetry, as well as both types of symmetry. We make the distinction because the techniques to exploit symmetry vary based on the type of symmetry. 

Furthermore, we can subdivide output symmetry into two more intersecting types—visible and invisible, where \textbf{visible output symmetry} is between indices that are present in the output tensor and \textbf{invisible output symmetry} is between indices that are not explicitly present in the output tensor, but are still involved in the computation. 

\begin{example}[Visible and Invisible Output Symmetry]
    \normalfont
    The assignment $B[i, j] = A[i, k] * A[j, k]$ exhibits visible output symmetry. Essentially, $B[i, j] = A[i, k] * A[j, k] = A[j, k] * A[i, k] = B[j, i]$. Thus, we know that B exhibits $\{\{i, j\}\}$ symmetry. Because the symmetry is preserved in the output, we refer to this symmetry as \textit{visible}.

    On the other hand, the assignment $B[i] = A[i, j] * A[i, k]$ exhibits invisible output symmetry. Let us rewrite the assignment with a temporary tensor $T$ as follows. 
    \begin{align*}
    T[i, j, k] &= A[i, j] * A[i, k] \\
    B[i] &= \sum_{j, k} T[i, j, k]
    \end{align*}

    Now the symmetry is more apparent: $T[i, j, k] = A[i, j] * A[i, k] = A[i, k] * A[i, j] = T[i, k, j]$. T (and thus B) exhibit $\{\{j, k\}\}$ symmetry. Because this symmetry is not seen in the output $B$, we refer to this symmetry as \textit{invisible}.
\end{example}

The two core strategies we have identified to exploit symmetry to make better use of memory and compute are (1) reusing canonical reads to save on bandwidth and (2) filtering redundant computations, which are dissected in more detail in the following subsections. \saman{How about storage savings?}

\subsection{Reusing Canonical Reads}
When input tensors are symmetric, we can restrict reads to the canonical triangle and use the same read to perform multiple computations for the output. This is critical for sparse inputs, since iteration over sparse inputs is expensive, especially if we must iterate in multiple directions at once, which is particularly relevant for iteration bound and memory bound kernels (e.g. SSYMV). The efficiency of these kernels is often limited by the rate at which data can be transferred from the memory to the processor (memory bandwidth) rather than the rate at which the processor can perform calculations (compute throughput). 

\begin{figure}
\begin{minipage}{0.5\linewidth}
\begin{lstlisting}[numbers=none, xleftmargin=0pt, basicstyle=\small]
for j=_, i=_
    y[i] += A[i, j] * x[j]       
\end{lstlisting}
\end{minipage}%
\begin{minipage}{0.5\linewidth}
\begin{lstlisting}[numbers=none, xleftmargin=0pt, basicstyle=\small]
for j=_, i=_
    if i < j
        a = A[i, j]
        y[i] += a * x[j]
        y[j] += a * x[i]
    if i == j
        y[i] += A[i, j] * x[j]
\end{lstlisting}
\end{minipage}
\caption{On left, a naive SSYMV. On right, SSYMV that accesses only canonical triangle \textit{and} reuses memory reads.}\label{fig:spmv_code}
\end{figure}

Let us take a look at what reusing canonical reads algorithmically entails for the sparse symmetric matrix-vector multiply (SSYMV) kernel given by $y[i] = A[i, j] * x[j]$. The optimization in Figure \ref{fig:spmv_code} limits accesses of the symmetric tensor to the canonical triangle and uses reads that are not on the diagonal for two assignments and reads that are on the diagonal for one assignment. Note that $i$ and $j$ are permuted in the second assignment and this makes up for not covering the iteration space where $i > j$. 

As the number of axes of symmetry increase, the complexity of the symmetry-optimized kernel increases, but so do the optimization opportunities. For instance, suppose that $A$ in the mode-1 TTM kernel \cite{kolda_tensor_2009} given by $C[i, j, l] = A[k, j, l] * B[k, i]$ is fully symmetry. The resulting kernel from restricting accesses of $A$ to the canonical triangle and performing all necessary updates to the output tensor $C$ is given by Listing \ref{lst:ttm}. 

\begin{lstlisting}[label={lst:ttm}, caption={TTM kernel that accesses only the canonical triangle of A.}]
for l=_, i=_, k=_, j=_
    if j <= k && k <= l
        if j < k && k < l
            A = A[j, k, l]
            C[i, j, l] += A * B[k, i]
            C[i, j, k] += A * B[l, i]
            C[i, k, l] += A * B[j, i]
            C[i, k, j] += A * B[l, i]
            C[i, l, k] += A * B[j, i]
            C[i, l, j] += A * B[k, i]
        if j == k && k != l
            A = A[j, k, l]
            C[i, j, l] += A * B[k, i]
            C[i, j, k] += A * B[l, i]
            C[i, l, k] += A * B[j, i]     
        if j != k && k == l  
            A = A[j, k, l]
            C[i, j, l] += A * B[k, i]
            C[i, k, l] += A * B[j, i]
            C[i, k, j] += A * B[l, i]
        if j == k && k == l
            C[i, j, l] += A[j, k, l] * B[k, i]
\end{lstlisting} 

The monotonically increasing condition on line 2 of Listing \ref{lst:ttm} enforces that we only iterate over the canonical triangle of symmetric tensor $A$. With three axes of symmetry, there are more diagonals to consider as the number of equivalence groups increase: i.e. the diagonals represented by equivalence groups $\{(j = k)\}$, $\{(k = l)\}$, $\{(j = l)\}$, and $\{(j = k = l)\}$. We handle each of these diagonals separately in Listing \ref{lst:ttm}, with the exception of $\{(j = l)\}$ because our overarching monotonically increasing condition ensures that if $j = l$, then we are overlapping the diagonal represented by $\{(j = k = l)\}$, which we already handle.

Given $n$ axes of symmetry, upon restricting a kernel to access only $\frac{1}{n!}$ of a tensor, we need to perform $n!$ assignments in each iteration to write to all the triangles of the output tensor in the case where none of the $n$ indices are equivalent. However, if $m$ indices are equivalent to each other (e.g. we read an element on a diagonal of the symmetric tensor) then we only perform $\frac{n!}{m!}$ assignments to avoid duplicate assignments. In other words, we perform the same number of assignments as unique permutations of the indices per iteration to make up for the fact that we are only covering $\frac{1}{n!}$ of the iteration space. The simplest solution to symmetrize code and handle these edge cases is to define every possible combination of equivalent indices and specify each assignment to the output, then optimize those statements.

\subsection{Optimizing Multiple Triangular Assignments} 
The symmetrization process results in multiple assignments being performed with one read of the symmetric tensors. Explicitly representing multiple triangular assignments makes plain the redundancies of symmetry and allows us to easily optimize or filter them. 

\subsubsection{Visible Output Symmetry}
Visible output symmetry involves indices that \textit{are} used to index the output tensor. In the presence of visible output symmetry, we can restrict our kernel to compute the values comprising only the canonical triangle of the output. Afterwards, we can perform an extra post-processing step that consists of copying the canonical triangle of the output to the other triangles. 

For example, let us consider the first block of the symmetrized TTM kernel given in Listing \ref{lst:ttm} that performs the assignments using coordinates of $A$ in the canonical triangle that are not on a diagonal. We reorder the assignments to make the pattern from output symmetry more obvious in Listing \ref{lst:ttm_output_symmetry_before}. Swapping the second and third indices in the output tensor on the left-hand side lends an equivalent right-hand side for each expression. As depicted in Listing \ref{lst:ttm_output_symmetry_after}, we can exploit the output symmetry by only writing to the canonical triangle of the output tensor (i.e. if we index \texttt{C} as \texttt{C[i, j, l]}, then only where \texttt{j <= l}), which reduces the number of computations that are done by a factor of $2$. Then, we can copy the values from the canonical triangles to the other triangles of the output tensor in a separate loop nest (lines 7-9 of Listing \ref{lst:ttm_output_symmetry_after}), thus completing the remaining assignments. 

\begin{lstlisting}[label={lst:ttm_output_symmetry_before}, caption={Before exploiting output symmetry in the conditional block of the TTM kernel that handles non-diagonal coordinates of A.}]
for l=_, j=_, k=_, i=_
    if j <= k && k <= l
        A = A[j, k, l]
        C[i, (*@\hl{j, l}@*)] += A * B[k, i]
        C[i, (*@\hl{l, j}@*)] += A * B[k, i]
        C[i, (*@\hl{j, k}@*)] += A * B[l, i]
        C[i, (*@\hl{k, j}@*)] += A * B[l, i]
        C[i, (*@\hl{k, l}@*)] += A * B[j, i]
        C[i, (*@\hl{l, k}@*)] += A * B[j, i]
\end{lstlisting}

\begin{lstlisting}[label={lst:ttm_output_symmetry_after}, caption={After exploiting output symmetry in the conditional block of the TTM kernel that handles non-diagonal coordinates of A.}]
for l=_, j=_, k=_, i=_
    if j <= k && k <= l
        A = A[j, k, l]
        C[i, j, l] += A * B[k, i]
        C[i, j, k] += A * B[l, i]
        C[i, k, l] += A * B[j, i]
for l=_, j=_, i=_
    if j > l
        C[i, j, l] = C[i, l, j]
\end{lstlisting}

In general, if $n$ indices are in the same part of a partition representing the visible symmetry of the output tensor, then we can reduce the number of operations by a factor of $n!$.

\subsubsection{Invisible Output Symmetry}
While visible output symmetry results in equivalent assignments to multiple locations, invisible output symmetry results in equivalent assignments to the same locations. We optimize redundant computation by replacing $k$ additions with equivalent right-hand sides with a single addition that multiples the right-hand side by scalar $k$. 

SYPRD is given by $y = x[i] * A[i, j] * x[j]$ where A is symmetric. SYPRD exemplifies invisible output symmetry because the output is a scalar (and thus any output symmetry must be with indices that are not present in the output). If we permute $i, j$, then we obtain an equivalent assignment. 
\[
    y = x[i] * A[i, j] * x[j] = x[j] * A[j, i] * x[i]
\]

Thus, instead of performing both non-diagonal assignments  
in Listing \ref{lst:syprd_output_symmetry_before} (lines 5-6), we can optimize by only performing one assignment but multiplying it by a factor of 2, as depicted in Listing \ref{lst:syprd_output_symmetry_before} (line 3). Note that this does not apply to the block that accesses the diagonal entries of $A$ because $i$ and $j$ are equivalent and thus there is only one assignment. 

\begin{lstlisting}[caption={SYPRD before exploiting output symmetry.}, label={lst:syprd_output_symmetry_before}, ]
for j=_, i=_
    if i <= j
        if i < j
            A = A[i, j]
            y[] += x[i] * A * x[j]
            y[] += x[j] * A * x[i]
        if i == j
            y[] += x[i] * A[i, j] * x[j]
\end{lstlisting}  

\begin{lstlisting}[caption={SYPRD after exploiting output symmetry}, label={lst:syprd_output_symmetry_after}, ]
for j=_, i=_
    if i < j
        y[] += 2 * x[i] * A[i, j] * x[j]
    if i == j
        y[] += x[i] * A[i, j] * x[j]
\end{lstlisting}   

Invisible output symmetry often presents itself when there are multiple of the same operands in an assignment. Using the same process depicted in the prior section, we may need to swap around a few indices in the blocks accounting for the diagonals to make the invisible output symmetry more apparent. This normalization makes it easier to pinpoint when assignments are equivalent.

If $n$ indices are in the same part of a partition representing the invisible symmetry of the output tensor, then we can reduce the number of operations by a factor of $n!$.

\section{Symmetric Compiler Methodology}
\label{sec:symmetric_compiler} 
Given an assignment and a map of input tensors that are known to be symmetric and the partitions that represent their symmetries, to take advantage of symmetry, we need to first generate a kernel that reuses memory reads and then, filter the resulting redundant computations. For simple assignments, it is easy to do this by hand, but as the number of indices involved in a symmetry group, the dimensionality of the tensors, and the number of tensors in the assignment increase, writing a symmetric kernel becomes less intuitive and more akin to a trial-and-error process. In this section, we propose a mechanical, generalizable system to generate a symmetry-exploiting kernel that is applicable to any tensor assignment and which can be replicated in any compiler. 

We divide this system in two phases to reflect the two core strategies of first capitalizing on memory bandwidth and then compute throughput. The first phase is \textit{symmetrization} and consists of generating code to read only the canonical triangle(s) of the symmetric tensor(s). The second phase is \textit{optimization} and consists of applying various transforms to reduce the number of memory accesses and operations that are performed.

\subsection{Symmetrization}
The process of symmetrization involves adding the appropriate control structures to limit the iteration space to the canonical triangles of the symmetric input tensors and determining which additional assignments will need to be made and and under what conditions to ensure that all the appropriate updates to the output tensor are performed.

We will use set $S_T$ to represent the equivalent permutations of a fully or partially symmetric tensor $T$. If $T$ is fully symmetric, $S_T$ is the set of all permutations of $\{1, ..., n\}$. If $T$ is partially symmetric with partition $\Pi$, $S_T$ is the set of all permutations $\sigma$ of $\{1, ..., n\}$ which only permute elements within their parts in $\Pi$.

Given an assignment $$O[i_1, ..., i_n] = T_1[i_{1,1}, ..., i_{1,n_1}] \otimes ... \otimes T_m[i_{m,1}, ..., i_{m,n_m}],$$ let $\Pi_i$ be the partition that defines the symmetry of $T_i$. Furthermore, we represent the symmetry groups as $S_{T_1}, ..., S_{T_m}$ and $S_O$. 

To represent and easily distinguish which diagonal of a tensor we are accessing, we introduce the notion of \textit{equivalence groups}—a term that we have formulated to represent the tensor generalizations of diagonals. 

\begin{definition}[Equivalence Group]
\normalfont
Given a set of indices $I$, we define \textit{equivalence group} $E$ to represent a partition $\Pi$ of indices $i\in I$ where for each part $\pi \in \Pi$, $i_n$ = $i_m$ for all $n, m \in \pi.$  
\end{definition}

We define the notation symmetry group $S_T | E$ to represent the unique permutations of a tensor's indices given a particular equivalence group $E$.

\begin{definition}[Unique Symmetry Group]
\normalfont
Let $S_T | E$ represent the \textit{unique symmetry group}, which given an equivalence group $E$, consists of $S_T | E = \{ \pi \in S_n \mid \forall i, j \in \{1, 2, ..., n\}$, if $i, j$ are both in the same subset of $E$, then $\pi(i) < \pi(j) \}.$
\end{definition}

The four stages below delineate the process to systematically generate a symmetrized kernel for this assignment. We assume that in addition to the assignment itself, the client has also provided the partitions $\Pi_i$ for each input tensor $T_i$ as well as the loop order (i.e. the order in which they will be looping through the indices in the assignment).  

\begin{enumerate}
    \item \textbf{Identify Symmetry}: First, we determine the set of permutable indices $P$, which is given by $$P = \bigcup_{i=1}^m \left( \bigcup \left\{ \pi \in \Pi_i \, \middle| \, |\pi| > 2 \right\} \right)$$ and includes all indices in the tensor assignment that are in a symmetry group with more than one index. Note that this step overapproximates symmetry—for instance, if we have $\{\{1, 2\}, \{3, 4\}\}$ symmetry in a tensor, we obtain $P = \{1, 2, 3, 4\}$. 

    \item \textbf{Restrict Iteration Space}: We establish an ordering $p_1, ..., p_n$ of the permutable indices in $P$ such that accessing any tensor $T_i$ at entries where $p_1, ..., p_n$ are monotonically increasing (i.e. $p_1 \leq ... \leq p_n$) will only access the canonical triangle of all symmetric tensors. This ordering is a topological sort of the dependence graph between canonical indices and always exists.
    
    \item \textbf{Define Assignments}: For each equivalence group $E$ that can be constructed from $P$ and satisfies the monotonically increasing condition established in step (2), we determine the unique symmetry group $S_P | E$ where $S_P$ consists of all the permutations of $P$. Then we can apply each permutation $\sigma \in S_P | E$ to the original assignment to generate all the assignments that need to be performed if the equivalence relationships defined by $E$ are satisfied.

    \item \textbf{Normalize Assignments}: Lastly, we normalize all assignments to make it easier to identify equivalent assignments or patterns across assignments during the optimization process.  There are many ways to rewrite an expression and yield an equivalent result; namely, indices in a symmetric group of a symmetric tensor can be permuted and operands involved in commutative operations can be commuted. Standardizing tensor assignments can make it easier to programmatically identify equivalent assignments and distinguish patterns across assignments. Thus, we define the notion of a \textit{normalized} assignment to be an assignment were (1) all tensors on the right-hand side have been ordered based on some predetermined sort order (e.g. alphabetical) and (2) for all symmetric tensors $T_i$ in the assignment, all indices in the same part of the partition $\Pi_i$ representing the symmetry of $T_i$ are ordered based on some predetermined sort order (e.g. to be concordant with the loop order). 
\end{enumerate}

The resulting symmetrized kernel from applying these steps is depicted via mathematical pseudocode in Figure \ref{fig:symmetrization_pseudocode}. We first enforce the monotonically increasing condition for the permutable indices (line 1) to restrict the iteration space to the canonical triangles of the symmetric tensors. We iterate through all possible equivalence groups of $P$ (line 3) and for each, determine the set of unique permutations of $P$ given the equivalence group (line 4). We apply each of these unique permutations to the initial assignment (line 6) to obtain all the assignments that are performed for the equivalence relationships represented by corresponding equivalence group. 

\begin{figure}[h]
    \fbox{\parbox{0.45\textwidth}{\begin{algorithmic}[1]
        \For {$i_1 = 1:\_$, $i_2=1:\_$, ...}  
        \If {$p_1 \leq ... \leq p_n$} 
            \ForAll {$E \text{ of } P$} 
                \State {Construct $S_P | E$}
                \ForAll {$\sigma \in S_P | E$}
                    \State $\left( O[i_1, ..., i_n] = T_1[i^1_1, ..., i^1_n] \otimes ... \otimes  / \right.$
                        \State \hspace{1.5em}$\left. T_m[i^m_1, ..., i^m_n]\right)[i \rightarrow \sigma(i)] $
                \EndFor 
            \EndFor   
        \EndIf 
        \EndFor
    \end{algorithmic}}}
    \caption{Pseudocode for Symmetrized Kernel}
    \label{fig:symmetrization_pseudocode}
\end{figure}

We can furthermore unroll the loops from lines 5-7 and lines 3-8 in Figure \ref{fig:symmetrization_pseudocode} to generate a more efficient kernel. Additionally, note that each equivalence group is exclusive (i.e. a coordinate only satisfies one of the equivalence groups), so when we do unroll the loops, the conditional blocks that are generated are exclusive.

\subsection{Optimization}
\label{subsec:optimization}
\begin{figure}
    \centering
    \includegraphics[width=\linewidth]{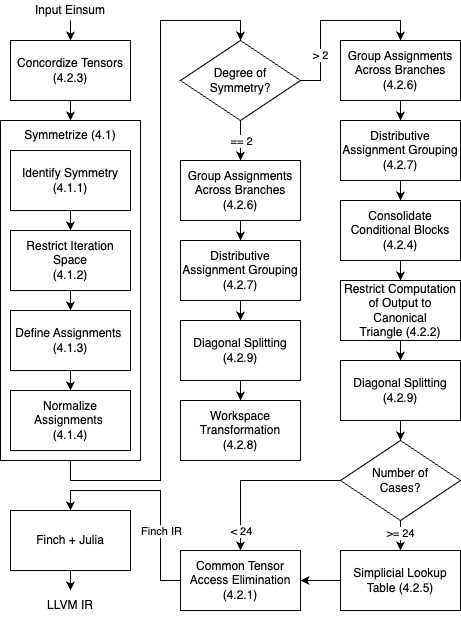}
    \caption{Symmetric Compiler Flow}
    \label{fig:compiler_flow}
\end{figure}

After symmetrizing the kernel such that it accesses only the canonical
triangle(s) of the symmetric tensor(s), we shift to applying various transforms
to reduce the number of computations performed. These transforms are the
building blocks for filtering redundant code. Several of these transformations may be familiar, but these transformations are
performed at the level of sparse tensor computation in Finch IR before Finch
lowers to Julia and then LLVM IR. Therefore, we cannot rely on the Julia or LLVM compilers
to perform these optimizations. The \textit{simplicial lookup tables} and \textit{diagonal splitting}
are novel, as is the overall composition of these transforms with the goal of
exploiting symmetry. The compiler is summarized in Figure \ref{fig:compiler_flow}. 

\subsubsection{Common Tensor Access Elimination}

We replace repeated reads of the same element in a tensor with a single constant value. In particular, after normalizing the symmetrized kernel, all accesses to a fully symmetric tensor will be equivalent in each iteration of a loop. For a fully symmetric tensor of order $n$, this will entail reducing memory reads by a factor of $n!$. Accesses to other tensors might also be repeated and thus, can also be consolidated. This step is crucial for the Finch compiler because it understands each tensor access as a separate iterator, and multiple redundant accesses would lead to intersecting multiple iterators in the loop. Thus, this step must be applied before running Finch.

\noindent\begin{minipage}[t]{0.50\linewidth}
\begin{lstlisting}[linewidth=\linewidth, backgroundcolor=\color{white}, numbers=none, basicstyle=\small, xleftmargin=0pt]
y[i] += A[i, j] * x[j]
y[j] += A[i, j] * x[i]
\end{lstlisting}
\end{minipage}%
\noindent\begin{minipage}[t]{0.50\linewidth}
\begin{lstlisting}[linewidth=\linewidth, backgroundcolor=\color{white}, numbers=none, basicstyle=\small, xleftmargin=0pt]
temp = A[i, j]
y[i] += temp * x[j]
y[j] += temp * x[i]
\end{lstlisting}
\end{minipage}

\subsubsection{Restrict Computation of Output to Canonical Triangle}

Identify assignments with equivalent right-hand sides that update symmetric entries of the output tensor (i.e. coordinates with particular indices swapped) in the same conditional block. In this case, replace the symmetric assignments with just one assignment to the canonical coordinate of the output tensor. We also mark the indices across which the output tensor will need to be replicated. After the kernel is computed, the canonical triangle of the output tensor can be replicated to the noncanonical triangles if needed. We chose to do the replication in a separate loop as the main loop may access the same output location multiple times. By keeping these loops separate, we avoid repeating the out-of-order access required to replicate the output tensor.

\noindent\begin{minipage}[t]{0.50\linewidth}
\begin{lstlisting}[linewidth=\linewidth, backgroundcolor=\color{white}, numbers=none, basicstyle=\small, xleftmargin=0pt]
for j=_, i=_
    if i <= j
        y[i, j] += A[i, j] * x[j]
        y[j, i] += A[i, j] * x[j]
\end{lstlisting}
\end{minipage}%
\noindent\begin{minipage}[t]{0.50\linewidth}
\begin{lstlisting}[linewidth=\linewidth, backgroundcolor=\color{white}, numbers=none, basicstyle=\small, xleftmargin=0pt]
for j=_, i=_
    if i <= j
        y[i, j] += A[i, j] * x[j]
for j=_, i=_
    if i > j
        y[i, j] = y[j, i]
\end{lstlisting}
\end{minipage}

\subsubsection{Concordize Tensors}

Transpose tensors to make the iteration of indices concordant \cite{ahrens_looplets_2023}.
A program is concordant when the order of indices in each tensor access match the order in which loops are nested around it.
If necessary, transpose the tensor \textit{and} reorder the loops to make iteration concordant.
This step is critical for sparse tensors, as we can only iterate over hierarchical sparse formats with a concordant traversal.
Concordant traversal is faster for dense tensors as well. Given a particular loop order and set of tensors, we prioritize transposing the tensors with fewer dimensions (and/or modifying the loop order if necessary) over those with more dimensions for efficiency.  

\noindent\begin{minipage}[t]{0.45\linewidth}
\begin{lstlisting}[linewidth=\linewidth, backgroundcolor=\color{white}, numbers=none, basicstyle=\small, xleftmargin=0pt]
for j=_, k=_, i=_
    C[i, j] += A[i, k] * B[k, j]
    C[k, j] += A[i, k] * B[i, j]
\end{lstlisting}
\end{minipage}%
\noindent\begin{minipage}[t]{0.55\linewidth}
\begin{lstlisting}[linewidth=\linewidth, backgroundcolor=\color{white}, numbers=none, basicstyle=\small, xleftmargin=0pt]
for k=_, i=_, j=_
    C_T[j, i] += A[i, k] * B_T[j, k]
    C_T[j, k] += A[i, k] * B_T[j, i]
\end{lstlisting}
\end{minipage}

\subsubsection{Consolidate Conditional Blocks}

Identify conditional blocks containing equivalent assignments and replace them with a single conditional block with an if-condition that is the union of the if-conditions of each of the conditional blocks. This transform improves the readability of the generated kernel and also prevents unnecessary specialization of cases during Finch compilation. 

\noindent\begin{minipage}[t]{0.50\linewidth}
\begin{lstlisting}[linewidth=\linewidth, backgroundcolor=\color{white}, numbers=none, basicstyle=\small, xleftmargin=0pt]
if i == j
    y[i] += A[i, j] * x[j]
if i < j
    y[i] += A[i, j] * x[j]
\end{lstlisting}
\end{minipage}%
\noindent\begin{minipage}[t]{0.50\linewidth}
\begin{lstlisting}[linewidth=\linewidth, backgroundcolor=\color{white}, numbers=none, basicstyle=\small, xleftmargin=0pt]
if (i == j) || (i < j)
     y[i] += A[i, j] * x[j]
\end{lstlisting}
\end{minipage}

\subsubsection{Simplicial Lookup Table}
Given multiple conditional blocks with the same assignments but with different constant factors, we combine them into a single block and generate a lookup table that is used to determine the constant factor. We index into the lookup table using some product of primes based on which indices are equivalent.  

\noindent\begin{minipage}[t]{0.5\linewidth}
\begin{lstlisting}[linewidth=\linewidth, backgroundcolor=\color{white}, numbers=none, basicstyle=\footnotesize, xleftmargin=0pt]
if (i != k) && (k != l)
    C[l, j] += 2 * A[i, k, l] * B[i, j]
    C[k, j] += 2 * A[i, k, l] * B[l, j]
    C[i, j] += 2 * A[i, k, l] * B[k, j]
if ((i != k) && (k == l)) 
            || ((i == k) && (k != l)) 
    C[l, j] += A[i, k, l] * B[i, j]
    C[k, j] += A[i, k, l] * B[l, j]
    C[i, j] += A[i, k, l] * B[k, j]
if (i == k) && (k == l) 
    C[l, j] += A[i, k, l] * B[i, j]
\end{lstlisting}
\end{minipage}%
\noindent\begin{minipage}[t]{0.5\linewidth}
\begin{lstlisting}[linewidth=\linewidth, backgroundcolor=\color{white}, numbers=none, basicstyle=\footnotesize, xleftmargin=0pt]
lookup_table = [2, 0, 1, 1, 0, 0, 1/3]
idx = 2 * (i == k) + 3 * (k == l) + 1
factor = lookup_table[idx]

C[l, j] += factor * A[i, k, l] * B[i, j]
C[k, j] += factor * A[i, k, l] * B[l, j]
C[i, j] += factor * A[i, k, l] * B[k, j]
\end{lstlisting}
\end{minipage}

\subsubsection{Group Assignments Across Branches}

Many of the same assignments are performed in different branches in the code generated from the symmetrization process. Restructure and reorganize the generated code such that each assignment is called only once. This particular transform is beneficial when the total number of unique assignments (after applying the previous transforms) is less than the number of conditional blocks and we only apply it when this is the case; it also improves the readability of the generated kernel and prevents unnecessary specialization of cases during compilation.

\noindent\begin{minipage}[t]{0.50\linewidth}
\begin{lstlisting}[linewidth=\linewidth, backgroundcolor=\color{white}, numbers=none, basicstyle=\small, xleftmargin=0pt]
if i < j
    y[i] += A[i, j] * x[j]
    y[j] += A[i, j] * x[i]
if i == j
    y[i] += A[i, j] * x[j]
\end{lstlisting}
\end{minipage}%
\noindent\begin{minipage}[t]{0.50\linewidth}
\begin{lstlisting}[linewidth=\linewidth, backgroundcolor=\color{white}, numbers=none, basicstyle=\small, xleftmargin=0pt]
if i < j || i == j
    y[i] += A[i, j] * x[j]
if i < j
    y[i] += A[i, j] * x[i]
\end{lstlisting}
\end{minipage}

\subsubsection{Distributive Assignment Grouping} 

Replace $N$ equivalent additions in a conditional block with a single addition that multiples the right-hand side by $N$.

\noindent\begin{minipage}[t]{0.50\linewidth}
\begin{lstlisting}[linewidth=\linewidth, backgroundcolor=\color{white}, numbers=none, basicstyle=\small, xleftmargin=0pt]
y[i] += A[i, j] * x[j]
y[i] += A[i, j] * x[j]
\end{lstlisting}
\end{minipage}%
\noindent\begin{minipage}[t]{0.50\linewidth}
\begin{lstlisting}[linewidth=\linewidth, backgroundcolor=\color{white}, numbers=none, basicstyle=\small, xleftmargin=0pt]
y[i] += 2 * A[i, j] * x[j]
\end{lstlisting}
\end{minipage}

\subsubsection{Workspace Transformation}

Replace a write to the output tensor in an assignment with a write to a temporary variable defined just inside the innermost loop $L$ that iterates through an index used to access the output tensor in the assignment. Accumulate updates in this temporary variable and write back the sum to the output tensor just at the end of this loop. This is worthwhile to do when there are more for loops inside $L$.

\noindent\begin{minipage}[t]{0.50\linewidth}
\begin{lstlisting}[linewidth=\linewidth, backgroundcolor=\color{white}, numbers=none, basicstyle=\small, xleftmargin=0pt]
for j=_, i=_
    y[i] += A[i, j] * x[j]
    y[j] += A[i, j] * x[i]
\end{lstlisting}
\end{minipage}%
\noindent\begin{minipage}[t]{0.50\linewidth}
\begin{lstlisting}[linewidth=\linewidth, backgroundcolor=\color{white}, numbers=none, basicstyle=\small, xleftmargin=0pt]
for j=_
    temp = 0
    for i=_
        y[i] += A[i, j] * x[j]
        temp += A[i, j] * x[i]
    y[j] += temp
\end{lstlisting}
\end{minipage}

\subsubsection{Diagonal Splitting}

Moving specific conditional \linebreak[4] blocks into a separate loop nest.  Because non-diagonal values form the bulk of the values in a tensor, we can think of assignments that involve the diagonal entries of a symmetric tensor as an edge case and compute them separately. In particular, we can move the conditional blocks involving non-diagonal entries in a separate loop nest. 

\noindent\begin{minipage}[t]{0.50\linewidth}
\begin{lstlisting}[linewidth=\linewidth, backgroundcolor=\color{white}, numbers=none, basicstyle=\small, xleftmargin=0pt]
for j=_, i=_
    if i != j
        y[i] += A[i, j] * x[j]
    if i == j
        y[i] += A[i, j] * x[j]
\end{lstlisting}
\end{minipage}%
\noindent\begin{minipage}[t]{0.50\linewidth}
\begin{lstlisting}[linewidth=\linewidth, backgroundcolor=\color{white}, numbers=none, basicstyle=\small, xleftmargin=0pt]
for j=_, i=_
    if i != j
        y[i] += A[i, j] * x[j]
for j=_, i=_
    if i == j
        y[i] += A[i, j] * x[j]
\end{lstlisting}
\end{minipage}

\subsection{MTTKRP Demonstration}

\begin{figure}[h]
    \scriptsize
    \parbox{\textwidth}{\begin{algorithmic}[1]
        \For {$j = 1:\_$, $l = 1:\_$, $k = 1:\_$, $i = 1:\_$}  
        \If {$i \leq k \leq l$} 
            \If {$E = \{(i), (k), (l)\}$} 
                % \ForAll {$\sigma \in \{(1, 2, 3), (1, 3, 2), (2, 1, 3), (2, 3, 1), \newline (3, 1, 2), (3, 2, 1) \}$}
                \ForAll {$\sigma \in \{(1,2,3),(1,3,2),(2,1,3),(2,3,1),(3,1,2),(3,2,1) \}$}
                    \State $(i, k, l) = \sigma((i, k, l))$
                    \State $C[i, j] = A[i, k, l] * B[l, j] * B[k, j]$
                \EndFor 
            \EndIf 
            \If {$E = \{(i = k), (l)\}$} 
                \ForAll {$\sigma \in \{(1, 2, 3), (1, 3, 2), (3, 1, 2)\}$}
                     \State $(i, k, l) = \sigma((i, k, l))$
                    \State $C[i, j] = A[i, k, l] * B[l, j] * B[k, j]$
                \EndFor 
            \EndIf 
            \If {$E = \{(i), (k = l)\}$} 
                \ForAll {$\sigma \in \{(1, 2, 3), (2, 1, 3), (3, 1, 2)\}$}
                    \State $(i, k, l) = \sigma((i, k, l))$
                    \State $C[i, j] = A[i, k, l] * B[l, j] * B[k, j]$
                \EndFor 
            \EndIf 
            \If {$E = \{(i = k = l)\}$} 
                \ForAll {$\sigma \in \{(1, 2, 3)\}$}
                     \State $(i, k, l) = \sigma((i, k, l))$
                    \State $C[i, j] = A[i, k, l] * B[l, j] * B[k, j]$
                \EndFor 
            \EndIf 
        \EndIf 
        \EndFor
    \end{algorithmic}}
    \caption{MTTKRP Symmetrization: We construct the unique symmetry groups given each equivalence group.}
    \label{fig:mttkrp_symmetrization_3}
\end{figure}
Let us apply this technique to the MTTKRP kernel given by $C[i, j] = A[i, k, l] * B[l, j] * B[k, j]$. If A is fully-symmetric, the set of permutable indices is given by $P = \{i, k, l\}$ and we can establish ordering $i, k, l$—such that if these indices are monotonically increasing, we will only access the canonical triangle of $A$. The equivalence groups that can be constructed from $P$ and which satisfy the that $i \leq k \leq l$ are $\{(i), (k), (l)\}$, $\{(i = k), (l)\}$, $\{(i), (k = l)\}$, and $\{(i = k = l)\}$.

Next, we determine the unique symmetry group $S_P | E$ for each equivalence group $E$. For instance, for equivalence group $ \{(i = k), (l)\}$, $S_P | E = \{(1, 2, 3), (1, 3, 2), (3, 1, 2)\}$. Thus, the pseudocode in Figure \ref{fig:symmetrization_pseudocode} expands to Figure \ref{fig:mttkrp_symmetrization_3}.
The normalized equivalent is given by Listing \ref{lst:normalized_mttkrp}.

\begin{lstlisting}[caption={Normalized Symmetric MTTKRP Kernel}, label={lst:normalized_mttkrp}]
function mttkrp(C, A, B)
    for l=_, j=_, k=_, i=_
        if i <= k && k <= l
            if i != k && k != l
                C[i, j] += A[i, k, l] * B[k, j] * B[l, j]
                C[i, j] += A[i, k, l] * B[k, j] * B[l, j]
                C[k, j] += A[i, k, l] * B[i, j] * B[l, j]
                C[k, j] += A[i, k, l] * B[i, j] * B[l, j]
                C[l, j] += A[i, k, l] * B[i, j] * B[k, j]
                C[l, j] += A[i, k, l] * B[i, j] * B[k, j]
            if i == k && k != l
                C[i, j] += A[i, k, l] * B[k, j] * B[l, j]
                C[i, j] += A[i, k, l] * B[k, j] * B[l, j]
                C[l, j] += A[i, k, l] * B[i, j] * B[k, j]
            if i != k && k == l
                C[i, j] += A[i, k, l] * B[k, j] * B[l, j]
                C[k, j] += A[i, k, l] * B[i, j] * B[l, j]
                C[k, j] += A[i, k, l] * B[i, j] * B[l, j]
            if i == k && k == l
                C[i, j] += A[i, k, l] * B[k, j] * B[l, j]
\end{lstlisting}

After performing common tensor access elimination, distributive assignment grouping, consolidating conditional blocks, diagonal splitting, and lastly, concordizing tensors, we obtain the code given by Listing \ref{lst:mttkrp_loop_nests}. 

\begin{lstlisting}[caption={MTTKRP: Separate Loop Nests}, label={lst:mttkrp_loop_nests}]
function mttkrp(C, A_nondiag, A_diag, B)
    for l=_, k=_, i=_, j=_
        A = A_nondiag[i, k, l], B_ji = B_T[j, i], B_jk = B_T[j, k], B_jl = B[j, l]
        if i <= k && k <= l
            if i != k && k != l
                C_T[j, i] += 2 * A * B_jk * B_jl
                C_T[j, k] += 2 * A * B_ji * B_jl
                C_T[j, l] += 2 * A * B_ji * B_jk
    for l=_, k=_, i=_, j=_
        A = A_diag[i, k, l], B_ji = B_T[j, i], B_jk = B_T[j, k], B_jl = B[j, l]
        if i <= k && k <= l
            if (i == k && k != l) || (i != k && k == l)
                C_T[j, i] += A * B_jk * B_jl
                C_T[j, l] += A * B_ji * B_jk
                C_T[j, k] += A * B_ji * B_jl
            if i == k && k == l
                C_T[j, i] += A * B_jk * B_jl
\end{lstlisting}

\section{Evaluation}
\subsection{Implementation}
We implemented the SySTeC compiler in Julia and demonstrated that the performance of the generated kernels was competitive on the SSYMV, SYPRD, SSYRK, TTM, and MTTKRP operations when compared against the naive Finch implementation\cite{ahrens_finch_2024}, TACO \cite{kjolstad_tensor_2017}, symmetric MKL \cite{noauthor_developer_2024}, and SPLATT \cite{smith_splatt_2015}. The implementation is available on github \footnote{\url{https://github.com/radha-patel/symmetry-compiler}}.

Provided a single Finch assignment and a list of symmetric tensors, SySTeC outputs an executable kernel in Finch IR that exploits symmetry. SySTeC uses RewriteTools\cite{noauthor_rewritetoolsjl_2024}, the same rewriting package used by Finch \cite{ahrens_looplets_2023}, to define a set of simplification rules and identify specific control structures, einsums, and operations to which these rules are applied.

SySTeC generates code in two phases according to Section \ref{sec:symmetric_compiler}: in the \textit{symmetrization} phase, the compiler first generates an executable kernel that only accesses the canonical triangle. In the \textit{optimization} phase, the compiler performs transforms to reduce operation count. Each transform from Section \ref{subsec:optimization} has been mapped into a rewrite rule that is applied if applicable.

\subsection{Results}
All experiments were run on a single core of a 12-core 2-socket Intel Xeon CPU E5-2680 v3 running at 2.50GHz with 128GB of memory. We used v0.6.22 of the Finch library to implement the kernels and executed both the naive and optimized implementation generated by SySTeC. We compare all kernels to the column-major implementations in TACO, and additionally SSYMV to MKLSparse v1.1.0 and MTTKRP to SPLATT. We used Julia v1.10 to run the tests and all timings are the minimum of 10,000 runs or 5s of measurement, whichever happens first. The time to rearrange data before or after each kernel is not included in the timings, including transposition or replicating the output.

\begin{table}[h!]
    \centering
    \footnotesize % Reduce font size
    \setlength{\tabcolsep}{3pt} % Adjust column separation
    \renewcommand{\arraystretch}{0.8} % Adjust row separation
    \begin{minipage}{0.5\linewidth}
        \begin{tabular}{lll}
            Name & Dimension & Nonzeros\\
            \hline
            bayer02 & 13935 & 63679\\
            bayer10 & 13436 & 94926\\
            bcsstk35 & 30237 & 1450163\\
            coater2 & 9540 & 207308\\
            crystk02 & 13965 & 968583\\
            crystk03 & 24696 & 1751178\\
            ct20stif & 52329 & 2698463\\
            ex11 & 16614 & 1096948\\
            finan512 & 74752 & 596992\\
            gemat11 & 4929 & 33185\\
            goodwin & 7320 & 324784\\
            lhr10 & 10672 & 232633\\
            lnsp3937 & 3937 & 25407\\
            memplus & 17758 & 126150\\
            nasasrb & 54870 & 2677324\\
        \end{tabular}
    \end{minipage}%
    \begin{minipage}{0.5\linewidth}
        \begin{tabular}{lll}
            Name & Dimension & Nonzeros\\
            \hline
            olafu & 16146 & 1015156\\
            onetone2 & 36057 & 227628\\
            orani678 & 2529 & 90185\\
            raefsky3 & 21200 & 1488768\\
            raefsky4 & 19779 & 1328611\\
            rdist1 & 4134 & 94408\\
            rim & 22560 & 1014951\\
            saylr4 & 3564 & 22316\\
            sherman3 & 5005 & 20033\\
            sherman5 & 3312 & 20793\\
            shyy161 & 76480 & 329762\\
            venkat01 & 62424 & 1717792\\
            vibrobox & 12328 & 342828\\
            wang3 & 26064 & 177168\\
            wang4 & 26068 & 177196
        \end{tabular}
    \end{minipage}
    \caption{Matrix collection used in the experiments, taken from Vuduc et. al \cite{vuduc_performance_2002}.}
    \label{tab:matrixtable}
\end{table}

For the SSYMV, SYPRD, and SSYRK kernels, we evaluated with the matrix benchmark suite used by Vuduc et. al \cite{vuduc_performance_2002}, shown in Table~\ref{tab:matrixtable}. The asymmetric matrices in the suite were symmetrized by summing the transpose (i.e. $A+A^T$). To our knowledge, there does not exist a database of symmetric tensors so for the MTTKRP kernels, we generated uniformly distributed symmetric random sparse tensors of varying sizes and sparsities via an Erdős–Rényi distribution. The dense input matrices are also randomly generated. The numerical rank is the number of columns in the dense matrix, where applicable. 

In Figures \ref{fig:ssymv_performance}-\ref{fig:mttkrp_performance}, we normalize all results to naive Finch; the red line specifies the performance of naive Finch (our baseline) and the purple line the speedup we expect. 

\subsubsection{SSYMV}
The sparse symmetric matrix vector kernel is given by $y[i] = A[i, j] * x[j]$ where A is symmetric and CSC, and y and x are dense.

\begin{figure*}[h]
\centering{\includegraphics[alt={SSYMV Performance},width=\linewidth]{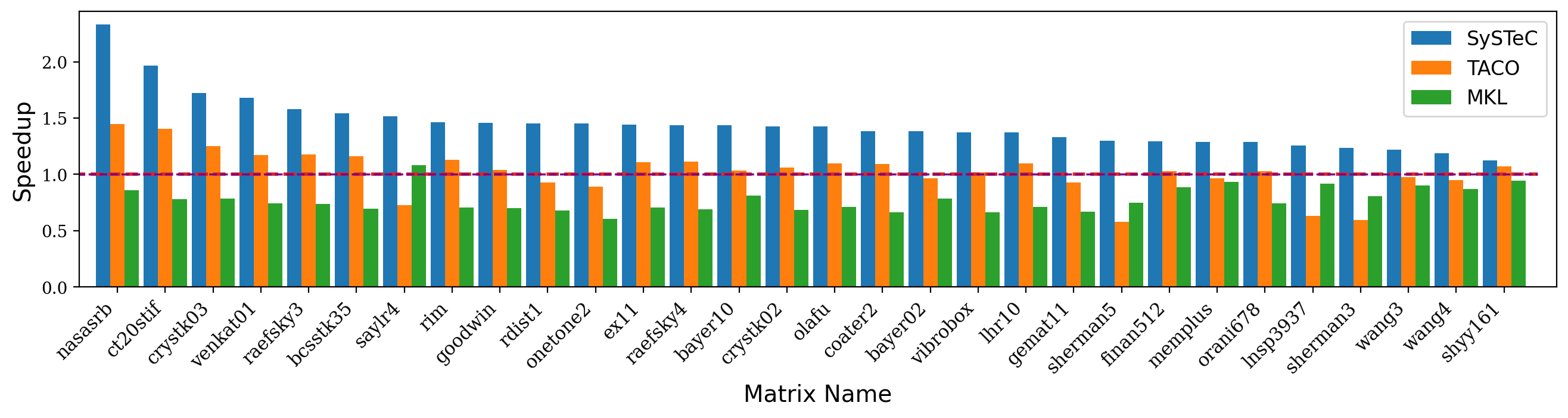}}%
\caption{SSYMV Performance}
\label{fig:ssymv_performance}
\end{figure*}

The optimized kernel accesses only $\frac{1}{2}$ of the values of A, but performs all of the computations. In cases where SSYMV is bandwidth bound, we can expect a speedup approaching 2x, however we don't expect any computational savings here. We find that SySTeC is 1.45, 1.45, and 1.90 times faster on average than the naive Finch implementation, TACO, and MKL's \texttt{mkl\_dcsrsymv}, respectively (Figure \ref{fig:ssymv_performance}). MKL was the only commercial sparse SYMV implementation for cpu we found, but is either not taking advantage of symmetry or not optimized for a single threaded case.
TACO may be faster than naive Finch because it emits simpler loop bounds for SPMV that are more amenable to compiler optimizations.

\subsubsection{Bellman-Ford Update}
\begin{figure*}[h]
\centering{\includegraphics[alt={Bellman Performance},width=\linewidth]{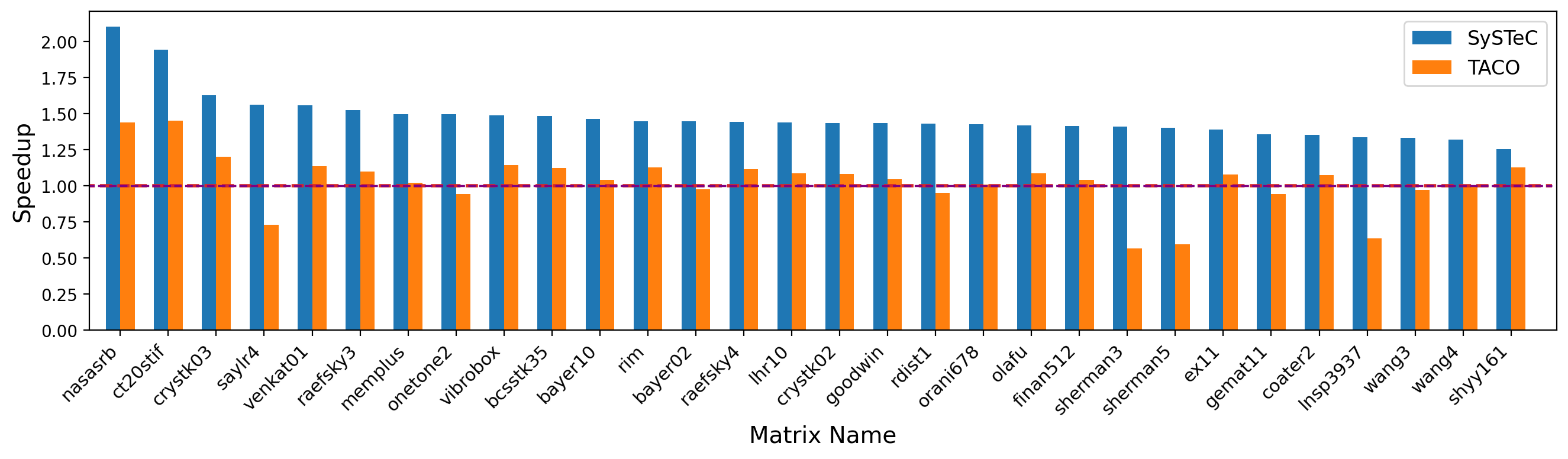}}%
\caption{Bellman-Ford Step Performance}
\label{fig:bellman_performance}
\end{figure*}

The sparse symmetric \linebreak[4] Bellman-Ford update kernel is given by $y[i] min= A[i, j] + d[j]$ where A is symmetric and CSC, and y and d are dense. Here, $d$ represents shortest path lengths after $k$ steps, $A$ represents the edge distances, and $y$ represents the shortest path lengths after $k+1$ steps.
This kernel is identical to SSYMV from a performance perspective, as seen in Figure \ref{fig:bellman_performance}, but is included to show that SySTeC can handle symmetrizing operations beyond $+$ and $*$.

\subsubsection{SYPRD}
The symmetric triple product kernel is given by $y[] = x[j] * A[i, j] * x[i]$ where A is symmetric and CSF and y and x are dense. 

\begin{figure*}[h]
\centering{\includegraphics[alt={SYPRD Performance},width=\linewidth]{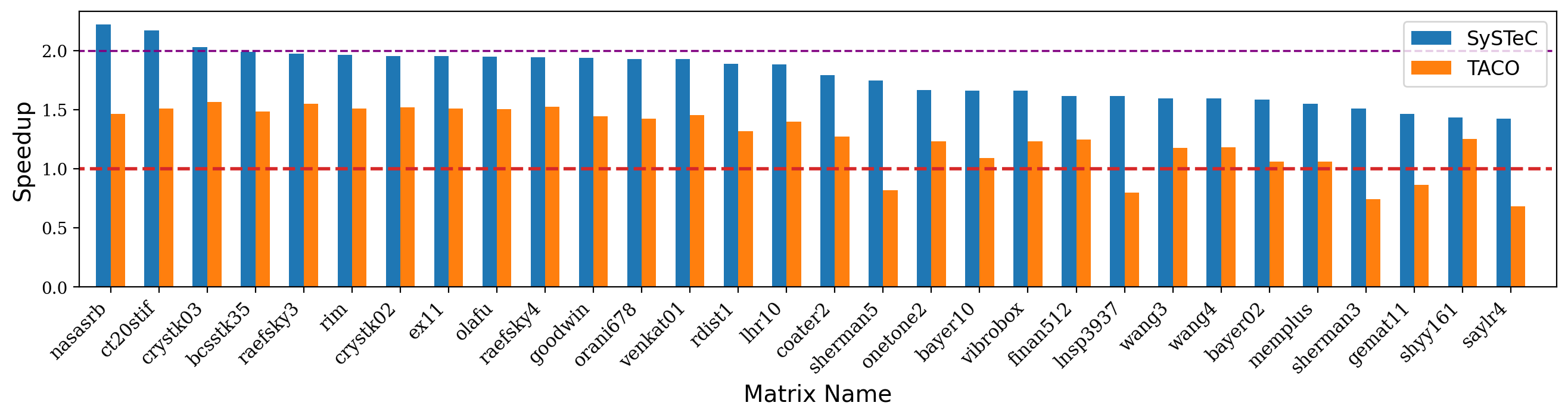}}%
\caption{SYPRD Performance}
\label{fig:syprd_performance}
\end{figure*}

The optimized kernel accesses $\frac{1}{2}$ of the values of A and performs $\frac{1}{2}$ of the computations because we have $\{\{i, j\}\}$ invisible symmetry in C. As $n$ grows, we can expect a speedup of 2x. We find that SySTeC is 1.79 and 1.46 times faster on average than naive Finch and TACO (Figure \ref{fig:syprd_performance}). We may not have achieved 2x speedup everywhere because the symmetric code needs to terminate the iteration through the sparse matrix early to restrict to the triangle, which complicates the exit condition of the loop.

\subsubsection{SSYRK}
The sparse symmetric rank-k update is given by $C[i, j] = A[i, k] * A[j, k]$ where A is \textit{not} symmetric, but by nature of the computation, C \textit{is} symmetric. A and C are both CSF.

\begin{figure*}[h]
    \centering{\includegraphics[alt={SSYRK Performance},width=\linewidth]{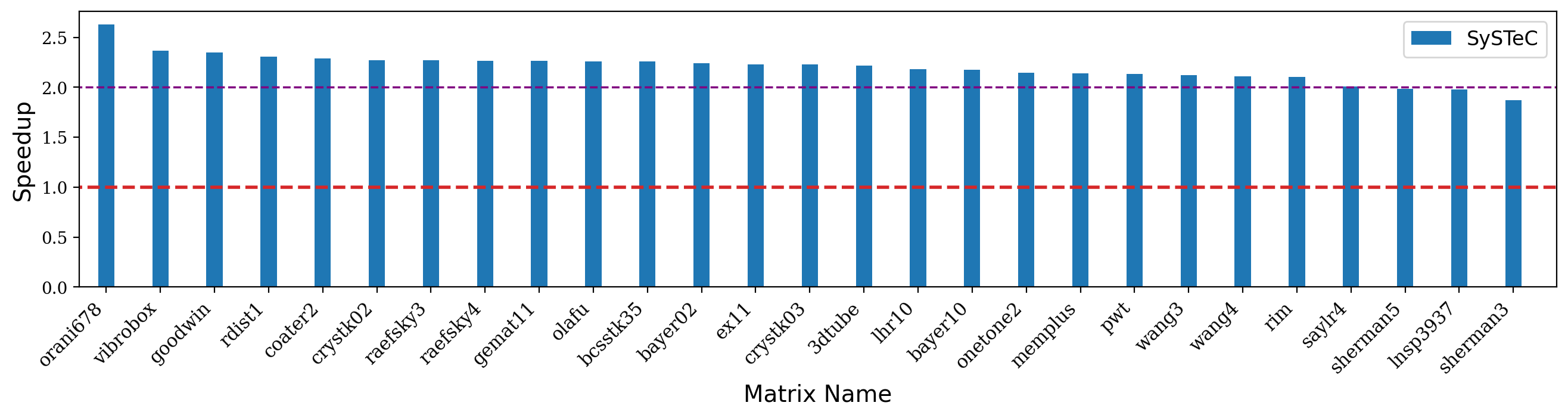}}%
    \caption{SSYRK Performance}
    \label{fig:ssyrk_performance}
\end{figure*}

The optimized kernel accesses all values of A because A is not symmetric, but performs only $\frac{1}{2}$ of the computations and writes to C because we exploit the $\{\{i, j\}\}$ output visible symmetry in C. Because SSYRK is compute-bound, we expect a speedup of 2x. We find that SySTeC is 2.20 times faster than naive Finch (Figure \ref{fig:ssyrk_performance}). TACO does not support the outer products implementation of SSYRK. We believe we exceed the expected speedup due to increased reuse of rows of A in the point of the triangle.

\subsubsection{TTM}
The tensor times matrix kernel is given by $C[i, j, l] = A[k, j, l] * B[k, i]$ where A is fully symmetric CSF, and B and C are dense. 

The optimized kernel accesses only $\frac{1}{6}$ of the values of A and performs $\frac{1}{2}$ of the computations (and hence writes $\frac{1}{2}$ of the values to C) because we take advantage of the $\{\{j, l\}\}$ symmetry in C. We can therefore expect a speedup of at least 2x. We find that SySTeC is 2.09 and 1.13 times faster than naive Finch and TACO, respectively, with high density and low numerical rank. SysTeC underperforms naive Finch for high numerical rank because the overhead of initializing the dense output outweighs the cost of the computation.

\begin{figure}[h]
\centering{\includegraphics[alt={TTM Performance},width=\linewidth]{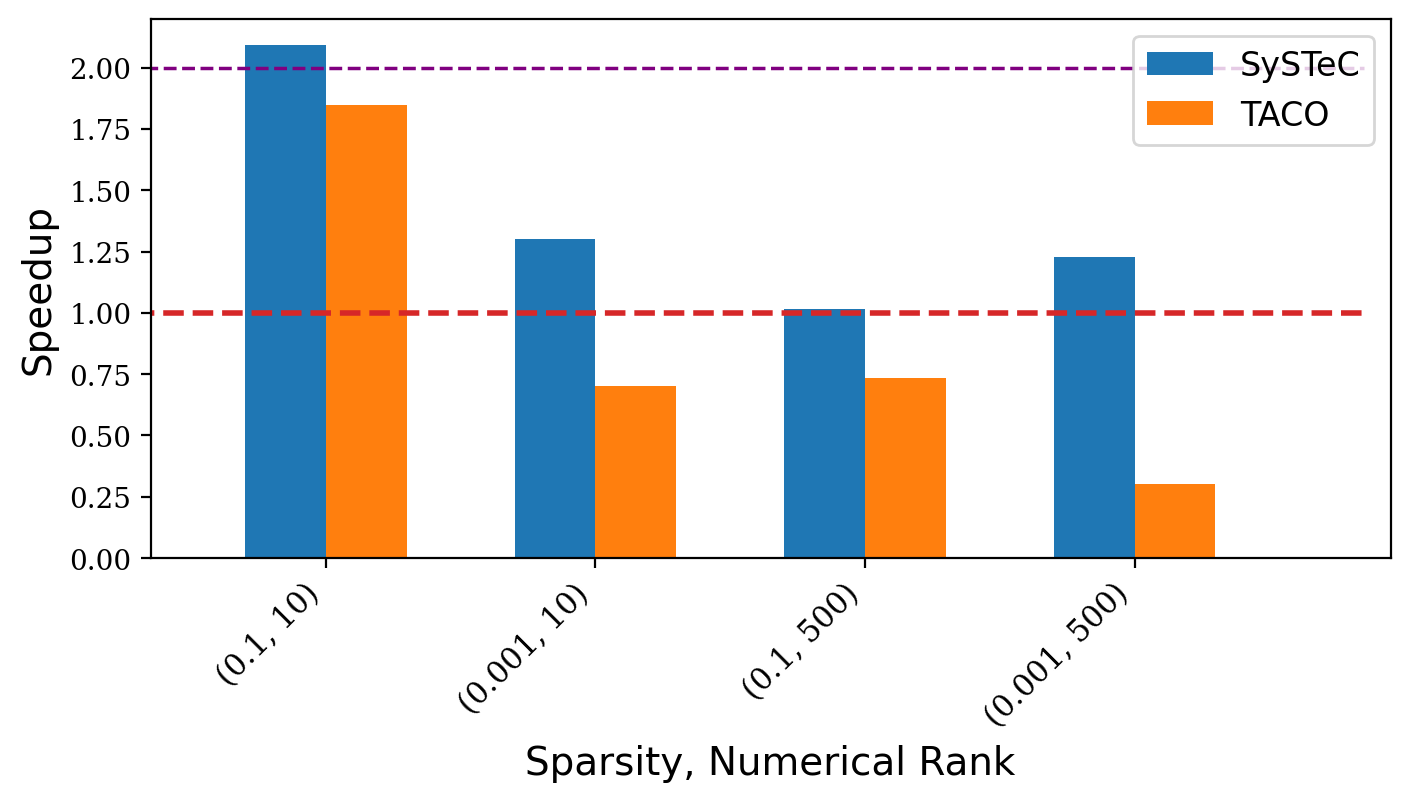}}%
\caption{TTM Performance}
\label{fig:ttm_performance}
\end{figure}

\subsubsection{MTTKRP}
The MTTKRP (matricized tensor times Khatri-Rao product) kernel is used in tensor factorization, namely to compute the CPD (Canonical Polyadic Decomposition) of a tensor \cite{kolda_tensor_2009}. Typically, the factorization of an $N$-dimensional
tensor requires $N$ MTTKRP kernels, each executed on a different transposition
of the tensor. When the tensor is symmetric, no transpose is required because all transpositions of the tensor are equivalent. 
While CPD typically involves separate factor matrices for each dimension, the symmetric CPD problem uses the same factor matrix for all dimensions.  This means that the update step is identical in all the modes, and the algorithm is more efficient \cite[Algorithm 2]{kofidis_best_2002}.

The assignments for the 3-, 4-, and 5-dimensional  kernels are given below where A is CSF and B and C are dense.

\begin{align*}
C[i, j] &= A[i, k, l] * B[k, j] * B[l, j] \\
C[i, j] &= A[i, k, l, m] * B[k, j] * B[l, j] * B[m, j] \\
C[i, j] &= A[i, k, l, m, n] * B[k, j] * B[l, j] * B[m, j] * B[n, j]
\end{align*}

\begin{figure*}[h]
\centering
\begin{minipage}{0.32\textwidth}
\centering
\includegraphics[width=\textwidth]{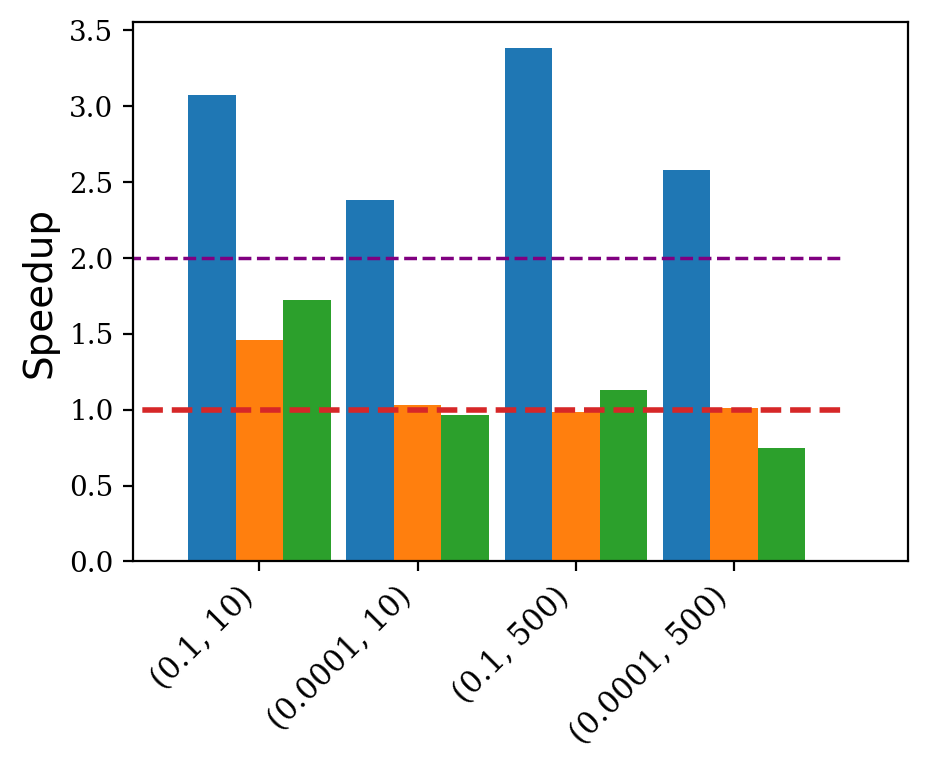}
% \caption{TODO}
% \label{fig:3d_mttkrp_performance}
\end{minipage}\hfill
\begin{minipage}{0.295\textwidth}
\centering
\includegraphics[width=\textwidth]{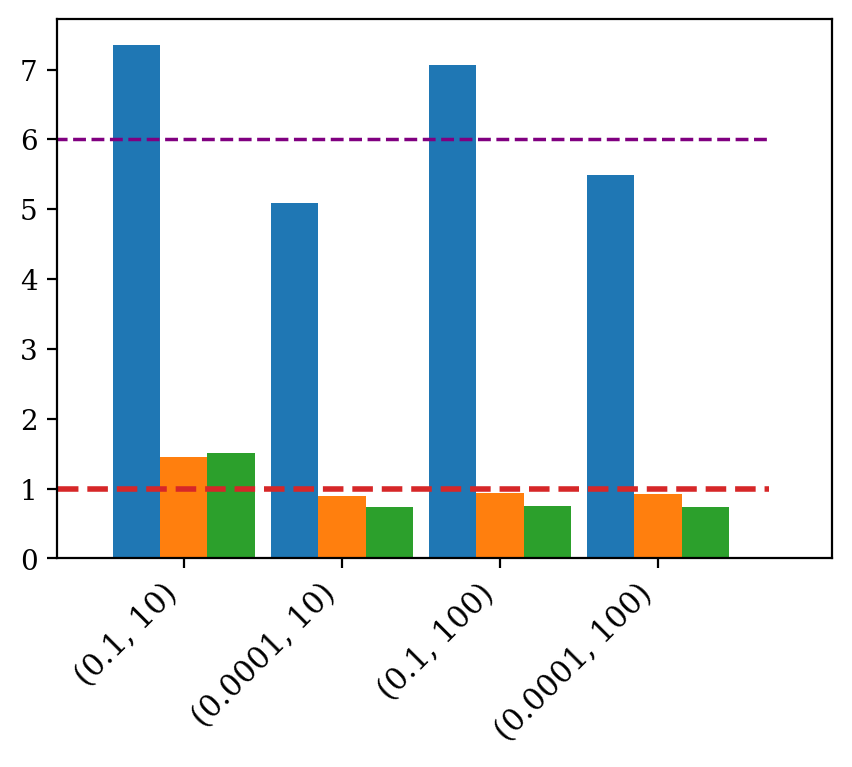}
% \caption{TODO}
% \label{fig:4d_mttkrp_performance}
\end{minipage}\hfill
\begin{minipage}{0.303\textwidth}
\centering
\includegraphics[width=\textwidth]{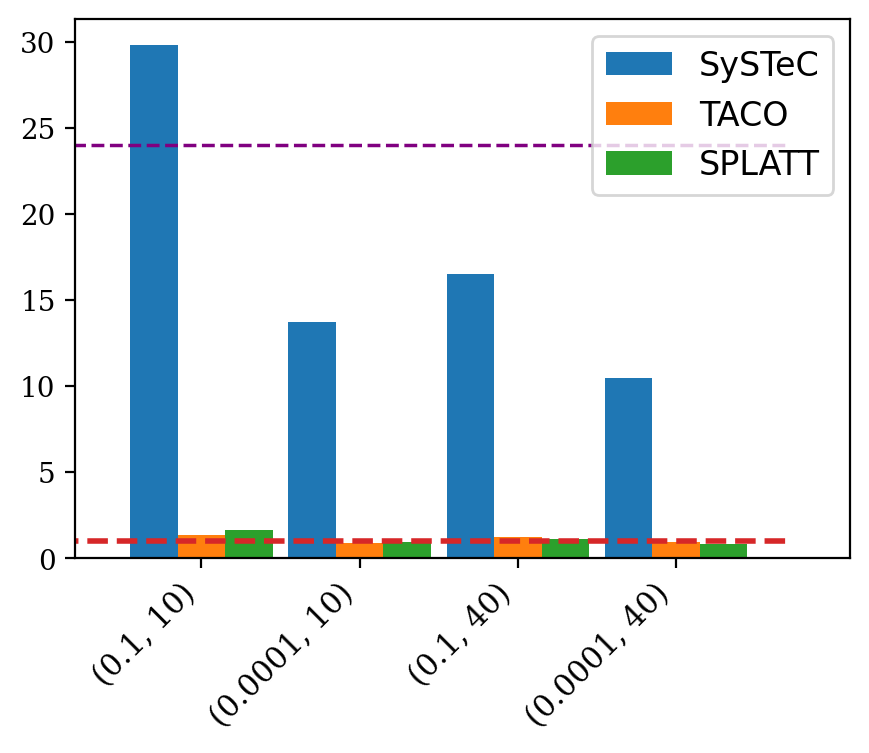}
% \caption{TODO}
% \label{fig:5d_mttkrp_performance}
\end{minipage}
\caption{3-, 4-, and 5-dimensional MTTKRP performance over varying sparsity and numerical rank}
\label{fig:mttkrp_performance}
\end{figure*}

The optimized kernels our compiler implementation generates for MTTKRP consist of two loop nests, one that handles the triangles and another to handle the diagonals to simplify control flow logic. For the 3D case, the optimized kernel accesses only $\frac{1}{6}$ of the values of A and performs $\frac{1}{2}$ of the computations because we have $\{\{k, l\}\}$ invisible symmetry in C. For the 4D case, the optimized kernel accesses only $\frac{1}{4!} = \frac{1}{24}$ of the values of A and performs $\frac{1}{3!} = \frac{1}{6}$ of the computations because we have $\{\{k, l, m\}\}$ invisible symmetry in C. Lastly, for the 5D case, the optimized kernel accesses only $\frac{1}{5!} = \frac{1}{120}$ of the values of A and performs $\frac{1}{4!} = \frac{1}{24}$ of the computations because we have $\{\{k, l, m, n\}\}$ invisible symmetry in C. Thus, we expect speedups of 2x, 6x, and 24x and obtain maximal speedups of 3.38, 7.35, and 29.8 times for 3-, 4-, and 5-dimensional MTTKRP, respectively, with SySTeC over naive Finch (Figure \ref{fig:mttkrp_performance}). We attribute the above-expected speedups over naive Finch to register reuse of the input tensors in the symmetric code.

\section{Related Work}
\label{sec:related_work}

Several tensor compilers support sparse tensor operations, but have no support for tensor symmetry, such as TACO \cite{kjolstad_tensor_2017}, SparseLNR \cite{dias_sparselnr_2022}, SPARTA \cite{liu_sparta_2021}, and MLIR \cite{bik_compiler_2022}. We therefore expect these systems would perform similarly to TACO in our evaluation.

Directly related techniques fall into three categories: Libraries which collect hand-specialized symmetric kernels, compilers which reduce symmetric problems to multiple asymmetric or hand-written kernels, and compilers which produce direct solutions to symmetric problems.

Many libraries contain at least the dense \linebreak[4] symmetry-specific functions specified in the BLAS \cite{anderson_lapack_1999}, such as ATLAS \cite{clint_whaley_automated_2001}, MKL \cite{noauthor_developer_2024}, and CuBLAS \cite{noauthor_cusparse_2024}. Of these, only MKL implements multiple sparse symmetric kernels. CuBLAS only handles sparse symmetry in SpMV. This reflects an implementation burden that further motivates our work.

Most of the work on symmetric compilers reduces symmetric problems to other kernels. The most notable of these is the Cyclops Tensor Framework (CTF). CTF reduces $N$-dimensional dense symmetric contractions to $N!$ separate triangular contractions, linearizes triangular indices which are preserved in the output, then dynamically loops over the remaining triangular indices and repeatedly calls matrix multiply, saving compute and storage \cite{solomonik_cyclops_2013, solomonik_sparse_2015}. However, this approach does not extend to kernels which are not contractions (such as MTTKRP) or do not use $+$ and $*$ (such as min-$+$ semiring multiplication \cite{buluc_design_2017}), and cannot benefit from within-kernel reuse of the redundant arguments that are produced by the problem reduction. This approach also requires transposing and reformatting arguments before running the kernel, which may be expensive in comparison to the cost of the kernel. It also only supports one sparse argument at a time (likely due to the complexity of sparse-sparse interactions in the triangular loops).
The OpMin system provides an operation optimization process that identifies the optimal ordering of tensors in a tensor contraction, but does not produce the code to compute the contraction \cite{lai_effective_2012}. 

%There are several existing solutions to optimize computation with symmetric tensors. The most common application of these solutions has been for the Coupled Cluster (CC) method, a numerical technique widely used for describing quantum many-body systems which involves hundreds to thousands of tensor contractions. Tensors in CC have a high-dimensional structure with permutational symmetry or skew-symmetry, which arises from the requirement that the wave-function for fermions (bosons) be antisymmetric (symmetric) under the interchange of particles \cite{solomonik_cyclops_2013}.

%Given the steep computational cost of quantum chemistry methods, a lot of effort has been put into designing algorithms that enable efficient use of computational resources. Overall, CC motivates tensor contraction algorithms that exploit symmetry in tensors, efficiently support contractions among tensors of diverse dimensions and shapes, and are suitable for long and repeated contraction sequences \cite{solomonik_cyclops_2013}. The goal remains to minimize communication (i.e. the number of words of data moved across network by any given processor) done to contract tensors.

Few compilers produce code that directly computes symmetric kernels.
Shi et. al. proposes to use an output-oriented loop structure which iterates through the unique inputs needed to compute the result, but the corresponding random access of the symmetric input resulted in poor performance \cite{shi_attempt_2021}.
STUR symbolically optimizes kernels based on the structure (triangular, banded, symmetric, etc.) of the arguments using a term rewriting approach, but cannot handle dynamic sparsity patterns, and is not as specialized to symmetry \cite{ghorbani_compiling_2023}.
Similarly, Spampinato proposes a polyhedral approach (sBLACs) to generating structured code, but does not address dynamic sparsity or even tensors of dynamic sizes \cite{spampinato_basic_2016}.
The ITensor library has several library routines for symmetric tensors within tensor networks, but these routines do not handle unstructured sparsity either \cite{fishman_itensor_2022}. 

Other works propose more specialized techniques for symmetric tensors which we do not attempt in this work. %
%Solomonik also proposes a specialized method to minimize operation count with the symv, syr2, syr2k, and symm kernels by taking advantage of multiplication on a ring by reformulating the kernels with a symmetric intermediate and introducing an additive inverse \cite{solomonik_cs294_lecture}. Furthermore, Cai et. al propose a new technique for computing the Matricized Khatri-Rao Product (MTTKRP) kernel that involves fusing the computations of the Khatri-Rao product and the multiplication of it with the matricized tensor, instead of completing one after the other \cite{cai2015optimization}.
Solomonik also proposes a Strassen-like algorithm to reduce operation count with the symv, syr2, syr2k, and symm kernels \cite{solomonik_fast_2021}, which is beyond the scope of this paper.
The Blocked Compact Symmetric format proposed by Schatz et. al for the TTM kernel breaks tensors into blocks, processing only canonical blocks of symmetric tensors \cite{schatz_exploiting_2013}. However, the implementation does not handle sparse tensors, and cannot optimize diagonal blocks.

\section{Future Work}

Several avenues for future research remain: 

\begin{enumerate}[topsep=8pt, parsep=2pt, partopsep=0pt]
    \item \textbf{Generalizing to More Types of Symmetry}: We can expand and adapt our methodology to encompass other forms of symmetry like antisymmetry, block symmetry, or cyclic symmetry that commonly arise in physics, mathematics, chemistry, and machine learning.
    \item \textbf{Exploring Parallelization Opportunities}:  Incorporating parallelization is important for increasing the practicality of our system, and are actively working towards that goal. In distributed memory settings, the block-cyclic distribution employed by Cyclops \cite{solomonik_cyclops_2013} would decompose the problem into several smaller, symmetric instances that we could process with SySTeC out of the box. In multicore settings, we plan to tag an outer loop as parallel. Of course, because iteration over the canonical triangle modifies different transpositions of the output at once, we will also need to make atomic updates to the output.
    \item \textbf{Fine-Tuning Symmetric optimizations}: Data formats tailored to
    symmetry that provide more efficient access when iterating over different
    transpositions of the canonical triangle could significantly improve
    performance. Symmetry-aware formats could also eliminate or simplify extra
    post-processing steps like replicating the canonical triangle of a tensor to
    the noncanonical triangles. Additionally, our implementation of Systec fully
    optimizes for every symmetry in the kernel, but we could develop approaches
    to partially optimize for symmetry only when it is beneficial.
\end{enumerate}

\pagebreak[4]

\section{Conclusion}
In this paper, we demonstrated a systematic approach to exploit symmetry in arbitrary tensor kernels. We identified core strategies to exploit symmetry in tensor kernels, including memory read reuse and redundant computation filtering. We also proposed a detailed compiler methodology for mechanically generating and optimizing symmetric code. This methodology involved two stages: first, symmetrizing the kernel such that we only access the canonical triangle of symmetric inputs, and secondly, applying a set of transforms to further optimize the code. We ultimately implemented this methodology in a Julia-based compiler and evaluated its performance on several common tensor kernels, showing significant speedups. 

This work provides a strong foundation for exploiting symmetry in tensor kernels.

\section{Data Availability Statement}
The data that support the findings of this study are openly available in Zenodo at \url{https://zenodo.org/records/13821280}. \cite{patel_systec_2024}.

\begin{acks}
    Intel and NSF PPoSS Grant CCF-2217064; DARPA PROWESS Award HR0011-23-C-0101; NSF SHF Grant CCF-2107244; DoE PSAAP Center DE-NA0003965; \newline DARPA SBIR HR001123C0139
\end{acks}

%\cleardoublepage
\newpage
\bibliographystyle{ACM-Reference-Format}
\bibliography{systec}
\clearpage

\appendix
%\cleardoublepage
\section{Artifact Appendix}

%%%%%%%%%%%%%%%%%%%%%%%%%%%%%%%%%%%%%%%%%%%%%%%%%%%%%%%%%%%%%%%%%%%%%
\subsection{Abstract}

This artifact provides all of the necessary tools to reproduce the program
compilation and charts presented in the paper, with the exception of SSYRK
(which takes too much time and memory), and MKL, which is too cumbersome to
install. The datasets given to TTM and MTTKRP are smaller here than in the
paper. We claim that all speedups should be within 1.5 of the expected, with few
exceptions. We note that there was a bug in the TACO TTM results in the submitted
version which has been fixed here, so the results may not
match the submitted version, but do match the camera ready.

\subsection{Artifact check-list (meta-information)}

\begin{itemize}
  \item {\bf Algorithm: } We present a new compilation algorithm, SySTeC, which
  accepts pointwise einsums with \linebreak[4] symmetry-annotated tensors and produces
  symmetry-optimized Finch tensor programs.
  \item {\bf Program: } We benchmark on the SSYMV, SYPRD, TTM, and MTTKRP
  Kernels. We compare against naive Finch, TACO, and SPLATT, all of which are
  included.
  \item {\bf Compilation: } We require Finch v0.6.32 and Julia v1.10.4 to compile.
  \item {\bf Binary: } We provide a Docker image \linebreak[4] \texttt{systec-artifact.tar} with all dependencies \linebreak[4] pre-installed, which requires ARM hardware to run as it was built for Apple M2. If different hardware is used, one can build the docker image for that hardware using the provided instructions.
  \item {\bf Data set: } Matrix datasets are downloaded automatically from \url{http://sparse.tamu.edu/}.
  \item {\bf Run-time environment: } The artifact requires a Unix System (Mac or Linux).
  \item {\bf Hardware: } We benchmark on an Intel Xeon CPU E5-2680 v3 running at 2.50GHz, and our benchmarks run well on Apple M2.
  \item {\bf Execution: } We assume that TACO and SPLATT will run in single-threaded mode.
  \item {\bf Metrics: } We measure runtime of all methods we test.
  \item {\bf Output: } The numerical output of each kernel is collected and
  validated against comparison methods in the script itself. The experiments
  themselves output json files in the same subdirectory they run in. The
  artifact includes instructions to plot the results.
  \item {\bf Experiments: } We claim that our speedup results should be within 1.5 of the expected, with few exceptions.
  \item {\bf How much disk space required (approximately)?: } 4GB
  \item {\bf How much time is needed to prepare workflow (approximately)?: } The artifact can be set up quickly with Docker, in around an hour. Building by hand also takes less than an hour, but it may take longer if your system needs special build flags.
  \item {\bf How much time is needed to complete experiments (approximately)?: } The experiments take around an hour to run.
  \item {\bf Publicly available?: } yes
  \item {\bf Code licenses (if publicly available)?: } MIT
  \item {\bf Workflow framework used?: } Docker
  \item {\bf Archived (provide DOI)?: } Zenodo \url{doi.org/10.5281/zenodo.13377504}
\end{itemize}

%%%%%%%%%%%%%%%%%%%%%%%%%%%%%%%%%%%%%%%%%%%%%%%%%%%%%%%%%%%%%%%%%%%%%
\subsection{Description}

\subsubsection{How delivered}
There are multiple sources:

Download from Zenodo: \url{doi.org/10.5281/zenodo.13377504}

Download from Github: \url{https://github.com/radha-patel/symmetry-benchmarks/tree/systec-cgo-artifact}

Disk space required: 4GB

\subsubsection{Hardware dependencies}

These experiments require at least 24GB of memory and 4GB disk space. A network
connection is required to download the matrix datasets.

We benchmark on an Intel Xeon E5-2695 v2 @ 2.40GHz, and our benchmarks run well on Apple M2.

\subsubsection{Software dependencies}

A Unix (Mac or Linux) system is required to run the artifact.  ARM hardware is
required to use the pre-built docker image included with the artifact, or one
can build their own docker image or build from source.

Dependencies which must be downloaded are documented in \texttt{Dockerfile}:
\begin{itemize}
    \item Julia v1.10.4
    \item coreutils
    \item cmake
    \item gcc
    \item g++
    \item python
    \item python3
    \item python3-pip
    \item python3-venv
    \item git
    \item libblas-dev
    \item liblapack-dev
    \item poetry
\end{itemize}

Dependencies which are built from source with \texttt{make deps}:
\begin{itemize}
\item TACO \url{https://github.com/tensor-compiler/taco/commit/1278503a1c859d557174a4ef2ae7a85295f39f69}
\item SPLATT \url{https://github.com/ShadenSmith/splatt/commit/6cb86283c1fbfddcc67c2564e025691de4f784cf}
\end{itemize}

Julia deps are also collected in a Project.toml file which can be installed with \verb|julia setup.jl| or just \verb|make env|.
\begin{itemize}
\item ArgParse v1.2.0
\item BenchmarkTools v1.5.0
\item DataStructures v0.18.20
\item Finch v0.6.32
\item JSON v0.21.4
\item MatrixDepot v1.0.13
\item SparseArrays v1.10.0
\item SySTeC v0.1.0 \url{https://github.com/radha-patel/SySTeC/commit/b0ec98927f0d2be01a61b48646cbadaa92040b0f}
\item TensorMarket v0.2.0
\end{itemize}

Python deps are collected in a pyproject.toml file which can be installed with \verb|poetry install --no-root| or just \verb|make env|.
\begin{itemize}
\item python = v3.9
\item matplotlib = v3.9.0
\item packaging = v24.0
\item numpy = v1.26.4
\end{itemize}

\subsubsection{Data sets}

All matrix datasets are automatically downloaded with a network connection from \url{http://sparse.tamu.edu/}

The tensor datasets are randomly generated by the benchmarks themselves.

%%%%%%%%%%%%%%%%%%%%%%%%%%%%%%%%%%%%%%%%%%%%%%%%%%%%%%%%%%%%%%%%%%%%%
\subsection{Installation}

\subsubsection{Docker}
Installation is greatly simplified if you are using the provided Docker image.
First load the image with
\begin{verbatim}
docker image load systec-artifact.tar
\end{verbatim}
or build it with
\begin{verbatim}
docker build -t systec-artifact .
\end{verbatim}

Then, you can launch a shell in the container named "evaluator-container" with
\begin{verbatim}
docker run -it --name evaluator systec-artifact
    /bin/bash
\end{verbatim}

Be sure to configure Docker to use enough memory (24GB) and disk space (100 GB) for the experiments.

\subsubsection{Manual}
Once the basic dependencies are installed on your system, you should be able to run \verb|make| to build all the necessary dependencies.

Should you need to customize the workflow, \verb|make deps| builds the
dependencies, \verb|make env| initializes Python and Julia Environments,
and \verb|make kernels| builds the benchmarking kernels.

%%%%%%%%%%%%%%%%%%%%%%%%%%%%%%%%%%%%%%%%%%%%%%%%%%%%%%%%%%%%%%%%%%%%%
\subsection{Experiment workflow}

There are three main phases to the experiments:

\subsubsection{Run SySTeC to compile the kernels}

From the toplevel directory, you can run the SySTeC compiler with

\begin{verbatim}
julia run_SySTeC.jl
\end{verbatim}

This will compile all the kernels required for the experiments and output them into the \verb|generated/| directory.
The script itself can be inspected and modified if you wish to compile different kernels and compare the output code.
The \linebreak[4] SySTeC compiler itself is contained in \linebreak[4] \verb|deps/SySTeC/src/SySTeC.jl|.

\subsubsection{Run the experiements}

You can run the experiments with
\begin{verbatim}
sh run_benchmarks.sh
\end{verbatim}
Benchmarks took us a little over an hour to run. The benchmarks may be run
individually with the individual commands in the \verb|run_benchmarks.sh|
script. Results will be generated in the toplevel directory. Corresponding reference results used in the
publication are available with the filename suffix \verb|_reference.json|.

\subsubsection{Plot the results}

You can plot the results with
\begin{verbatim}
poetry run python plot_results.py
\end{verbatim}

This will generate
plots for whichever experiments you have run and saved results for. The plots
will be saved in the \verb|charts/| directory. Corresponding reference results used in the
publication are available with the filename suffix \verb|_reference.json|.

%%%%%%%%%%%%%%%%%%%%%%%%%%%%%%%%%%%%%%%%%%%%%%%%%%%%%%%%%%%%%%%%%%%%%
\subsection{Evaluation and expected result}

You should expect to see similar results to the paper, notably SSYMV (Figure
\ref{fig:ssymv_performance}), SYPRD (Figure \ref{fig:syprd_performance}),
TTM (Figure \ref{fig:ttm_performance}), and MTTKRP (Figure \ref{fig:mttkrp_performance}).  We expect
that the more dramatic speedups will be seen in the MTTKRP benchmarks, as the
compiler is able to exploit the most symmetry in 5-dimensional kernels. Note
that we have reduced the size of the tensors in TTM and MTTKRP to keep the
runtime and storage manageable, so these results may demonstrate slightly less
speedup as more time is spent on diagonal edge cases.

If you are using our provided Docker image or using Docker to run the
experiments, you can view the charts by copying them to the host.
\begin{verbatim}
docker cp evaluator:symmetry-benchmarks/charts .
\end{verbatim}

%%%%%%%%%%%%%%%%%%%%%%%%%%%%%%%%%%%%%%%%%%%%%%%%%%%%%%%%%%%%%%%%%%%%%
\subsection{Experiment customization}

The evaluator may optionally customize the kernel given to SySTeC by
modifying the \verb|run_SySTeC.jl| script. The size of the tensors in our TTM
and MTTKRP experiments can be changed by modifying the scripts directly.

%%%%%%%%%%%%%%%%%%%%%%%%%%%%%%%%%%%%%%%%%%%%%%%%%%%%%%%%%%%%%%%%%%%%%
\subsection{Methodology}

Submission, reviewing and badging methodology:

\begin{itemize}
  \item \url{http://cTuning.org/ae/submission-20190109.html}
  \item \url{http://cTuning.org/ae/reviewing-20190109.html}
  \item \url{https://www.acm.org/publications/policies/artifact-review-badging}
\end{itemize}

\end{document}